\documentclass[twocolumn,superscriptaddress]{revtex4-2}
\usepackage{physics}
\usepackage{graphicx}
\usepackage{here}
\usepackage{amsmath}
\usepackage{amssymb}
\usepackage{bm}
\usepackage{wrapfig}
\usepackage{ascmac}
\usepackage{mathrsfs}
\usepackage{color}
\usepackage{comment}
\usepackage[normalem]{ulem}
\usepackage{amsthm}
\usepackage{mathtools}

\theoremstyle{plain}

\theoremstyle{definition}

\theoremstyle{remark}

\theoremstyle{plain}
\newtheorem*{thm*}{Theorem}
\newtheorem*{lem*}{Lemma}
\newtheorem*{prop*}{Proposition}
\newtheorem*{cor*}{Corollary}
\newtheorem*{conj*}{Conjecture}

\theoremstyle{definition}
\newtheorem*{ass*}{Assumption}
\newtheorem*{dfn*}{Definition}

\theoremstyle{remark}
\newtheorem*{rem*}{Remark}

\newcommand{\im}{{\rm i}}

\usepackage{comment}
\usepackage{hyperref}
\usepackage[normalem]{ulem}
\hypersetup{colorlinks=true,citecolor=blue,linkcolor=blue,urlcolor=blue}

\begin{document} 
\title{Quenching from superfluid to free bosons in two dimensions: entanglement, symmetries, and quantum Mpemba effect}
\author{Shion Yamashika}
\affiliation{Department of Physics, Chuo University, Bunkyo, Tokyo 112-8551, Japan} 
\author{Pasquale Calabrese}
\affiliation{SISSA and INFN, via Bonomea 265, 34136 Trieste, Italy}
\affiliation{International Centre for Theoretical Physics (ICTP), Strada Costiera 11,
34151 Trieste, Italy}
\author{Filiberto Ares}
\affiliation{SISSA and INFN, via Bonomea 265, 34136 Trieste, Italy}

\begin{abstract}
We study the non-equilibrium dynamics of bosons in a two-dimensional optical lattice after a sudden quench from the superfluid phase to the free-boson regime. 
The initial superfluid state is described approximately using both the Bogoliubov theory and the Gaussian variational principle. 
The subsequent time evolution remains Gaussian, and we compare the results from each approximation of the initial state by examining different aspects of the dynamics.
First, we analyze the entanglement entropy and observe that, in both cases, it increases linearly with time before reaching a saturation point. 
This behavior is attributed to the propagation of entangled pairs of quantum depletions in the superfluid state. 
Next, we explore the fate of particle-number symmetry, which is spontaneously broken in the superfluid phase. 
To do so, we use the entanglement asymmetry, a recently introduced observable that enables us to track symmetry breaking within a subsystem. 
We observe that its evolution varies qualitatively depending on the theory used to describe the initial state. 
However, in both cases, the symmetry remains broken and is never restored in the stationary state.
Finally, we assess the time it takes to reach the stationary state by evaluating the quantum fidelity between the stationary reduced density matrix and the time-evolved one. Interestingly, within the Gaussian variational principle, we find that an initial state further from the stationary state can relax more quickly than one closer to it, indicating the presence of the recently discovered quantum Mpemba effect. We derive the microscopic conditions necessary for this effect to occur and demonstrate that these conditions are never met in the Bogoliubov theory. 
\end{abstract}
\maketitle

\section{Introduction}
Non-equilibrium quantum systems have garnered significant attention during last years due to their rich and distinctive phenomenology, which is absent in equilibrium~\cite{pssv-11, ge-16, cem-16}. In particular, the non-equilibrium dynamics of isolated many-body quantum systems has emerged as a central problem, especially in the effort to bridge the gap between statistical and quantum physics. 
While isolated extended quantum systems evolve unitarily in time and, as such, do not inherently relax to a stationary configuration, the state of a portion of them does converge at large times into a statistical ensemble such as a Gibbs ensemble~\cite{deutsch-91, srednicki-94} or a generalized Gibbs ensemble when the system is integrable~\cite{rdyo-07, bs-08, cdeo-08, cef-12-2, fe-13, vr-16, ef-16}. 
Thus, understanding the mechanisms of relaxation of closed quantum systems is crucial for elucidating the microscopic origin of statistical mechanics. 
\par 
Quantum quenches are one of the most suitable protocols for that purpose: the system is initially prepared in the ground state of a specific Hamiltonian and then it is driven to non-equilibrium by suddenly changing a parameter of the Hamiltonian. This setup has been the subject of an intense research activity in the last years, in which the entanglement entropy has played a preeminent role as a probe of relaxation. Typically, the entanglement entropy increases linearly in time after the quench and, eventually, it converges to a finite value~\cite{cc-05} that corresponds to the entropy of the corresponding stationary statistical ensemble~\cite{dls-13, ac-17,c-18}. This behavior can be understood in terms of the excitation and the ballistic propagation of entangled pairs of quasiparticles~\cite{cc-05}. This picture has been proved in one-dimensional free~\cite{fc-08, ep-08, pe-09, nr-14, bkc-14, bfsed-16, hbmr-18, dvdr-24} and interacting integrable models~\cite{ac-17, ac-17-2, ac-17-3, ckc-14, mpc-19, mac-20, kb-21, kbp-21,c-20} as well as in two-dimensional free fermionic systems~\cite{yac-24, gjgb-24}. The quasiparticle picture also captures the quench dynamics of other quantities, see e.g.~\cite{cc-06, ac-19, ac-19-2, ctc-14, mac-22, dubail-17, ghkv-18, lsa-21, ggstlap-22, pbc-21, kfkr-20, bkl-22, rrc-24, ac-21, ca-22, hr-24}. In particular, R\'enyi entanglement entropies, which are directly measurable in experiments, are described by this picture in free systems, but not in the interacting integrable ones.  In that case, a more general unifying framework based on the duality of space and time, which also encompasses chaotic and non-integrable systems with no quasiparticles, has been recently introduced~\cite{bkalc-22, bcckr-23, bkccr-24}.
\par 
On the other hand, an aspect that has been usually overlooked is how fast the subsystem reaches the equilibrium after the quench. Very recently, this question has been addressed using broken symmetries as a proxy. The system is prepared in a non-equilibrium state that breaks certain internal symmetry, which is respected by the post-quench Hamiltonian. 
In one dimension, the symmetry is generically restored in the stationary state but, unexpectedly, for certain pairs of non-equilibrium initial states, the symmetry is earlier restored in the case in which it is initially more broken~\cite{amc-23}. This has been identified as a quantum version of the yet mysterious Mpemba effect --- the more a system is out of equilibrium, the faster it relaxes~\cite{mpemba, lasanta-17, lu-17, krhv-19, tyr-23, kb-20, kcb-22, bkc-21}.  To monitor the evolution of the symmetry in a subsystem, a novel observable dubbed entanglement asymmetry has been introduced~\cite{amc-23} (see~\cite{mnssob-24} for another recent approach to dynamical symmetry restoration). Using entanglement asymmetry, the mechanisms and conditions under which this effect occurs are well understood in terms of the quasiparticle picture for one-dimensional free fermionic chains~\cite{makc-24, carc-24} (also in the presence of dissipation and dephasing~\cite{cma-24, avm-24}), in interacting integrable systems~\cite{ bkccr-24, rylands-24, klobas-24, fcb-24, rvc-24}, and for free fermions in 2D lattices~\cite{yac-24-2}. 
The quantum Mpemba effect has also been found in holographic CFTs~\cite{bgs-24} as well as in non-integrable and chaotic systems~\cite{liu-24, tcdl-24, liu-24-2}, although the reasons and conditions for its occurrence in these cases are yet not completely clear. Other versions of the Mpemba effect in open quantum systems under different non-unitary dynamics have been simultaneously discovered~\cite{nf-19, kcl-22, cll-21, manikandan-21, ias-23, cth-23, cth-24, spc-24, mczg-24, bg-24, ne-24, longhi-24, wsw-24, befb-24}.
The quantum Mpemba effect has been also observed in experiments with an ion-trap quantum simulator, by directly measuring the entanglement asymmetry~\cite{joshi-24}; see also Refs.~\cite{shapira-24, zhang-24} for experimental evidence of other versions of the effect.
\par 
Thanks to remarkable advances in experimental techniques within the realm of quantum simulators~\cite{gb-17, hq-18, br-12}, the study of non-equilibrium many-body quantum systems has transitioned from purely theoretical to experimentally accessible. In particular, systems of bosons trapped in optical lattices, described by the Bose-Hubbard model~\cite{fisher-89, jaksch-98, sachdev}, are especially significant due to their high tunability and the broad spectrum of phenomena they can realize~\cite{bloch-08}.
An optical lattice is a periodic array of potential wells for neutral atoms, created using laser interference patterns. 
The elegance of this system lies in the precise control it offers over the depth and shape of the potentials, achieved by adjusting the intensity and angle of the lasers~\cite{greiner-02}. 
This tunability makes optical lattices an ideal platform for exploring non-equilibrium phenomena in isolated quantum systems.
For example, the quench dynamics of entanglement triggered by a sudden change in lattice depth, along with the subsequent relaxation of a subsystem towards a stationary state, have been experimentally analyzed in this setup~\cite{trotzky-11, islam-15, kaufman-16, cheneau-12, takasu-20}. 
\par 
In the one-dimensional case, there are analytic results on the time evolution of the entanglement entropy and correlation functions in quenches from the Mott-insulating limit to the strongly correlated~\cite{ykyt-23} and the free-boson~\cite{kkysytd-23} regimes. 
Instead, in two dimensions, particularly in the thermodynamic limit, there is still much to explore and understand. Recent numerical studies, such as those in Refs.~\cite{knqs-14, kd-22}, highlight these ongoing inquiries.
The Bose-Hubbard model is interacting and non-integrable~\cite{ch-82, kb-04} and, therefore, it is difficult to find analytic and numerical tools to study its non-equilibrium properties, especially in higher dimensions. 
\par 
In this work, we investigate the quench dynamics of bosons trapped in a two-dimensional optical lattice. In our case, the system is initially prepared in the superfluid state by a shallow lattice potential and then quenched into a free-boson system. To describe the initial superfluid state, we employ the standard Bogoliubov theory~\cite{bogoliubov-47} and the novel Gaussian variational principle, a systematic extension of the former introduced in Ref.~\cite{guaita-19}. Of course, numerous theoretical techniques have been developed to study the superfluid phase of the Bose-Hubbard model, see e.g.~\cite{bs-97, aa-02, dvods-03, sd-05, htab-08, kav-11, pekker-12}, but, in the Bogoliubov theory and the Gaussian variational principle, the superfluid state is approximated as a coherent squeezed Gaussian state. 
This allows us 
to apply the powerful tools of Gaussian states to derive analytically the time evolution of several quantities. We compute the R\'enyi entanglement entropies as well as the entanglement asymmetry of the particle-number symmetry, which is spontaneously broken in the superfluid state. We will see that this symmetry is not restored after the quench in a subsystem, even though the post-quench Hamiltonian respects it. Therefore, we cannot utilize it to investigate how fast the stationary state is reached and the appearance of the quantum Mpemba effect. Alternatively, we take the quantum fidelity between the time-evolved reduced density matrix and its corresponding stationary value. Our results reveal that the two theories for the superfluid state predict qualitatively distinct features in the time evolution of the subsystem. 
Indeed, the analysis of quantum fidelity indicates that the quantum Mpemba effect can emerge when using the Gaussian variational principle, whereas it is absent in the Bogoliubov theory.
We will see that quantum depletions (i.e., bosons ejected from the Bose-Einstein condensate by the repulsive interaction) play a crucial role in the quench dynamics of the system. 
The discrepancies observed between the quench dynamics predicted by the Bogoliubov theory and the Gaussian variational principle arise from differences in the mode occupancies of quantum depletions in the initial configuration. 
\par 
The organization of this paper is as follows: In Sec.~\ref{sec:model and setup}, we introduce the Bose-Hubbard model, which describes bosons in an optical lattice, and the quench protocol that we investigate. 
In Sec.~\ref{sec:effectiv theories}, we review the Bogoliubov theory and the Gaussian variational principle to describe the initial superfluid state. 
We study the dynamics of the entanglement entropy and of the entanglement asymmetry in Secs.~\ref{sec:entanglement entropy} and~\ref{sec:entanglement asymmetry}, respectively. 
In Sec.~\ref{sec:quantum fidelity}, we analyze the quantum fidelity of the subsystem with respect to the stationary state and determine the microscopic conditions for the quantum Mpemba effect to occur, showing that it is absent in the Bogoliubov theory. 
We finally summarize our results in Sec.~\ref{sec:conclusion}. 
We also include several appendices, where we describe in detail the derivation of certain formulas in the main text.

\section{Model and Setup}\label{sec:model and setup}

We study bosons trapped in a two-dimensional optical lattice. At zero temperature, the system is well-described by the Bose-Hubbard model~\cite{fisher-89, jaksch-98, sachdev}, 
\begin{align}
    H 
    = 
    -
    \frac{1}{2}
    \sum_{\bf \langle i,i'\rangle }
    (a_{\bf i}^\dag a_{\bf i'}+{\rm h.c.})
    + 
    \frac{U}{2}
    \sum_{\bf i}(a_{\bf i}^\dag)^2 a_{\bf i}^2 
    -\mu 
    \sum_{\bf i}a_{\bf i}^\dag a_{\bf i}, 
    \label{eq:Bose-Hubbard Hamiltonian}
\end{align}
where $\mathbf{i}=(i_x,i_y)$ is a vector identifying a site on the lattice, $U>0$ is the on-site repulsive interaction, $\mu$ is the chemical potential, $a_{\bf i}$ $(a_{\bf i}^\dag)$ is the annihilation (creation) operator of bosons at the $\mathbf{i}$-th site, and $\langle {\bf i,i'}\rangle$ stands for the summation over the nearest-neighbor sites. 
We denote as $L_x$ and $L_y$ the system length along the longitudinal and transverse directions, respectively. 
We impose periodic boundary conditions along both directions and, for simplicity, assume that $L_x$ and $L_y$ are even. 
\par 
The Bose-Hubbard model~\eqref{eq:Bose-Hubbard Hamiltonian} exhibits a quantum phase transition between the Mott-insulating and superfluid phases~\cite{fisher-89,oosten-01,greiner-02}. 
When the on-site interaction $U$ is dominant, the ground state is Mott-insulating, with bosons localized at each site due to the strong repulsive interactions.  
Meanwhile, when the hopping term is dominant, the ground state exhibits superfluidity, characterized by the majority of bosons being condensed in the zero-momentum mode. 
\par 
In the present work, we consider a system prepared in the ground state $\ket{{\rm SF}}$ deep in the superfluid phase of~\eqref{eq:Bose-Hubbard Hamiltonian} and perform a sudden global quantum quench to the free-boson system described by 
\begin{align}
    H_{\rm free}
    = 
    -\frac{1}{2}
    \sum_{\langle {\bf i,i'}\rangle}
    (a_{\bf i}^\dag a_{\bf i'}
    +{\rm h.c.})
    + 
    2\sum_{\bf i} a_{\bf i}^\dag a_{\bf i}. 
    \label{eq:H_free}
\end{align}
That is, at $t=0$, we instantaneously change the value of $U$ and $\mu$ in the Bose-Hubbard Hamiltonian~\eqref{eq:Bose-Hubbard Hamiltonian} from $U> 0$ and $\mu>-2$ to $U=0$ and $\mu=-2$. After the quench, the system evolves unitarily as 
\begin{equation}\label{eq:time_ev}
\ket{\Psi(t)}=e^{-\im t H_{\rm free}}\ket{{\rm SF}}.
\end{equation}

The free-boson Hamiltonian~\eqref{eq:H_free} is quadratic in the bosonic operators $a_{{\bf i}}$ and $a_{{\bf i}}^\dagger$, unlike the Bose-Hubbard model when $U\neq 0$. This will drastically simplify our analysis and will be crucial to derive the analytic results of this work. In fact, in contrast to the generic Bose-Hubbard Hamiltonian,  $H_{{\rm free}}$ can be exactly diagonalized by performing the Fourier transform 
\begin{align}
    a_{\bf k}
    = 
    \frac{1}{\sqrt{V}}
    \sum_{\bf i}
    e^{-\im {\bf k\cdot i}}
    a_{\bf i},
    \label{eq:Fourier transform}
\end{align}
where $V=L_xL_y$ is the volume of the whole system and $\mathbf{k}=(k_x,k_y)$, with $k_{x(y)}=2\pi n_{x(y)}/L_{x(y)}-\pi$ ($n_{x(y)}=0,1,...,L_{x(y)}-1$), 
are the quasimomenta. Applying Eq.~\eqref{eq:Fourier transform} in Eq.~\eqref{eq:H_free}, we find 
\begin{align}\label{eq:H_free_diag}
    H_{{\rm free}} 
    = 
    \sum_{\bf k}
    \xi_{\bf k}
    a_{\bf k}^\dag a_{\bf k}, 
\end{align}
with the single-particle dispersion relation $\xi_{\bf k}=2-\cos k_x-\cos k_y$. 
\par
\begin{figure}
    \centering
    \includegraphics[width=0.7\linewidth]{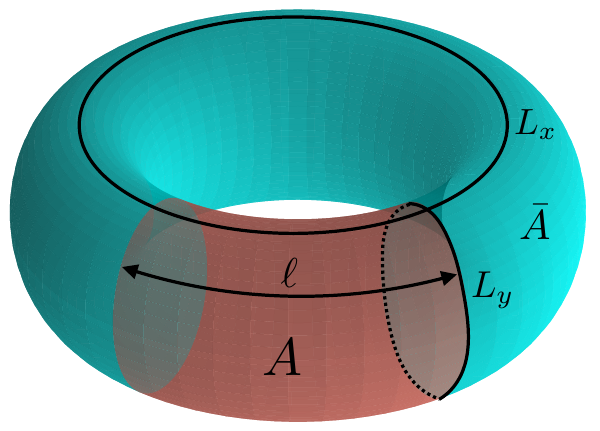}
    \caption{Schematic representation of the system under analysis. The red and blue regions represent subsystem $A$ and its complement $\bar A$, respectively.}
    \label{fig:torus}
\end{figure}

 
Since the time evolution~\eqref{eq:time_ev} in the quench protocol is unitary, the whole system will never relax even at large times.
Instead, we focus on a subsystem $A$ taken as the periodic stripe of width $\ell$ depicted in Fig.~\ref{fig:torus}. 
The state of the subsystem $A$ is fully described by the reduced density matrix 
\begin{align}
    \rho_A(t) 
    = 
    \Tr_{\bar A}(\ket{\Psi(t)}\bra{\Psi(t)}), 
    \label{eq:RDM}
\end{align}
where $\Tr_{\bar A}$ stands for the partial trace for $\bar A$, the complement of $A$. 
Although the whole system will never relax, in generic circumstances, the reduced density matrix $\rho_A(t)$ is expected to tend at large times to a stationary state determined by the initial configuration and the post-quench Hamiltonian~\eqref{eq:H_free}. 

\section{Effective theories for the superfluid state}\label{sec:effectiv theories}

In this section, we review the Bogoliubov theory and the Gaussian variational principle, which are the effective theories that we will employ to describe the initial state of our quench protocol, i.e., the ground state of the Bose-Hubbard model~\eqref{eq:Bose-Hubbard Hamiltonian} deep in the superfluid phase.

\subsection{Bogoliubov theory}
The Bogoliubov theory~\cite{bogoliubov-47} is the simplest way to characterize the superfluid state taking into account the quantum fluctuations. To introduce it, let us first rewrite the Bose-Hubbard model~\eqref{eq:Bose-Hubbard Hamiltonian} in the Fourier space~\eqref{eq:Fourier transform},
\begin{align}
    H 
    = 
    \sum_{\bf k} 
    \epsilon_{\bf k}a_{\bf k}^\dag a_{\bf k}
    + 
    \frac{U}{2V}
    \sum_{\bf k,k',q}
    a_{\bf k+q}^\dag 
    a_{\bf k'-q}^\dag 
    a_{\bf k}
    a_{\bf k'},
    \label{eq:Bose-Hubbard Fourier}
\end{align}
where 
\begin{align}
    \epsilon_{\bf k}
    = 
    -\sum_{\nu=x,y}\cos k_\nu -\mu
\end{align}
is the single-particle dispersion relation.  
Since most of the bosons are condensed in the ${\bf k=0}$ mode deep in the superfluid phase, we expect that the ground state can be well approximated by the coherent state, 
\begin{align} 
    \mathcal{D}(n_0)\ket{0},
    \label{eq:D|0>}
\end{align}
where $\ket{0}$ is the bosonic vacuum state annihilated by all $a_{\bf i}$ and 
\begin{align}
    \mathcal{D}(n_0)
    = 
    \exp(\sqrt{n_0}\sum_{\bf i}(a_{\bf i}^\dag -a_{\bf i}))
    \label{eq:displacement operator}
\end{align}
is the displacement operator with $n_0$ being the condensate fraction. In the coherent state~\eqref{eq:D|0>}, the superfluid order parameter takes a finite value,  
\begin{align}
    \bra{0}\mathcal{D}^\dag (n_0) 
    a_{\bf i}
    \mathcal{D}(n_0)\ket{0}
    =\sqrt{n_0}. 
\end{align}
This implies that the U(1) particle-number symmetry, which is preserved by the Bose-Hubbard Hamiltonian~\eqref{eq:Bose-Hubbard Hamiltonian}, is spontaneously broken in the superfluid phase. 
We choose $n_0$ such that the expectation value of the energy in the state~\eqref{eq:D|0>}, 
\begin{align}
    \bra{0}\mathcal{D}^\dag (n_0)
    H
    \mathcal{D}(n_0)\ket{0}
    = 
    V\qty(
    \epsilon_{\bf 0}n_0
    + 
    \frac{U}{2}n_0^2),
\end{align}
takes its minimum value. 
This can be done by choosing $n_0$ as 
\begin{align}
    n_0 
    = 
    -\frac{\epsilon_{\bf 0}}{U}
    = 
    \frac{2+\mu}{U}.
    \label{eq:n_0}
\end{align}
Since the condensate fraction should be positive, $n_0>0$, we assume $\mu>-2$ in the following. 
\par 
So far we have considered only the bosons inside the condensate. 
However, in addition, there exist bosons outside the condensate due to the on-site repulsive interaction when $U>0$. 
We will refer to such bosons as quantum depletions.
The annihilation operator of an excitation in the quantum depletion with the momentum $\mathbf{k}$ is defined as 
\begin{align}
     b_{\bf k}
    =
    \mathcal{D}(n_0)
    a_{\bf k}
    \mathcal{D}^\dag (n_0)
    = 
    a_{\bf k}-\delta_{\bf k,0}\sqrt{V n_0}.
\end{align}
Given that most of the bosons are in the superfluid, we can expect that quantum depletions are dilute and, therefore, the interaction between them can be neglected. This assumption allows us to approximate the Bose-Hubbard model~\eqref{eq:Bose-Hubbard Hamiltonian} to a quadratic Hamiltonian in terms of $b_{\bf k}$ and $b_{\bf k}^\dag$,  
\begin{align}
    H 
    &\simeq 
    H_{\rm Bog}= 
    \frac{1}{2}
    \sum_{\bf k}
    [(\epsilon_{\bf k}-2\epsilon_{\bf 0})b_{\bf k}^\dag 
    b_{\bf k}
    -\epsilon_{\bf 0}
    b_{\bf k}^\dag 
    b_{\bf -k}^\dag 
    +{\rm h.c.}],
    \label{eq:H_Bog}
\end{align}
by truncating the third and the fourth order terms in $b_{\bf k}$ and $b_{\bf k}^\dag$. The first-order terms in $b_{\bf k}$ vanish in Eq.~\eqref{eq:H_Bog} due to the constraint in Eq.~\eqref{eq:n_0}. 
\par 
As well known, the quadratic Hamiltonian~\eqref{eq:H_Bog} can be diagonalized via the Bogoliubov transform
\begin{gather}
    \eta_{\bf k}
    = 
    \cosh(\frac{\theta_{\bf k}}{2}) b_{\bf k}
    + 
    \sinh(\frac{\theta_{\bf k}}{2}) b_{\bf -k}^\dag, 
    \label{eq:Bogoliubov transform}
\end{gather}
where $\theta_{\bf k}$ is the Bogoliubov angle, which in the case of the Hamiltonian~\eqref{eq:H_Bog} is given by 
\begin{gather}
    \theta_{\bf k}
    = 
    \tanh^{-1}\qty(
    \frac{\epsilon_{\bf 0}}{\epsilon_{\bf k}-2\epsilon_{\bf 0}}). 
    \label{eq:Bogoliubov angle}
\end{gather}
Applying Eq.~\eqref{eq:Bogoliubov transform} in Eq.~\eqref{eq:H_Bog}, we eventually obtain 
\begin{align}
    H_{\rm Bog}
    = 
    \sum_{\bf k}
    \sqrt{(\epsilon_{\bf k}-2\epsilon_{\bf 0})^2-\epsilon_{\bf 0}^2}
    \qty(\eta_{\bf k}^\dag \eta_{\bf k}+\frac{1}{2}).
    \label{eq:H_Bog_diagonal}
\end{align}
Note that the Bogoliubov quasiparticles created and annihilated by $\eta_{\bf k}^\dagger$ and $\eta_{\bf k}$ are massless according to their dispersion relation in Eq.~\eqref{eq:H_Bog_diagonal}. 
This implies that they are the Nambu-Goldstone modes originating from the spontaneous breaking of the U(1) particle-number symmetry. 
\par 
From Eq.~\eqref{eq:H_Bog_diagonal}, we can see that the ground state deep in the superfluid phase $\ket{\rm SF}$ is the vacuum annihilated by the operators $\eta_{\bf k}$. 
To derive its explicit form, it is convenient to introduce the squeezing operator 
\begin{align}
    \mathcal{S}(\{\theta_{\bf k}\})
    = 
    \exp(\frac{1}{4}\sum_{\bf k}\theta_{\bf k}
    a_{\bf k}^\dag a_{\bf -k}^\dag -{\rm h.c.}). 
    \label{eq:squeezing operator}
\end{align}
Using the displacement operator~\eqref{eq:displacement operator} and the squeezing operator~\eqref{eq:squeezing operator}, we can relate $\eta_{\bf k}$ with $a_{\bf k}$ as 
\begin{align}
    \eta_{\bf k}
    = 
    \mathcal{D}(n_0)
    \mathcal{S}(\{\theta_{\bf k}\})
    a_{\bf k}
    \mathcal{S}^\dag(\{\theta_{\bf k}\})
    \mathcal{D}^\dag (n_0)
    . 
    \label{eq:Bogoliubov transform 2}
\end{align}
Finally, combining Eq.~\eqref{eq:Bogoliubov transform 2} and the conditions $\eta_{\bf k} \ket{\rm SF}=0~\forall{\bf k}$, we obtain the explicit form of $\ket{\rm SF}$, 
\begin{align}
    \ket{\rm SF}
    = 
    \mathcal{D}(n_0) 
    \mathcal{S}(\{\theta_{\bf k}\})
    \ket{0}.
    \label{eq:|SF> Bogoliubov}
\end{align}
According to Eq.~\eqref{eq:|SF> Bogoliubov}, quantum depletions with opposite momenta form condensed pairs in the state $\ket{\rm SF}$. 
This means that not only the condensate but also the quantum depletions contribute to the spontaneous breaking of the U(1) particle-number symmetry. We will discuss this in more detail in Sec.~\ref{sec:entanglement asymmetry}. 
\par 
We note that the superfluid state~\eqref{eq:|SF> Bogoliubov} is actually ill-defined because the Bogoliubov angle~\eqref{eq:Bogoliubov angle} diverges in the long-wavelength limit ${\bf k}\to {\bf 0}$. The standard way to avoid this problem is to force $\theta_{\bf 0}=0$ by hand.  
This choice corresponds to assuming that there are no quantum depletions with zero momentum. 
In other words, all the particles with zero momentum are in the condensate. 
Hereafter, we assume $\theta_{\bf 0}=0$ whenever we employ the Bogoliubov theory. 
\begin{figure}
    \raggedright
    \includegraphics[width=0.49\textwidth]{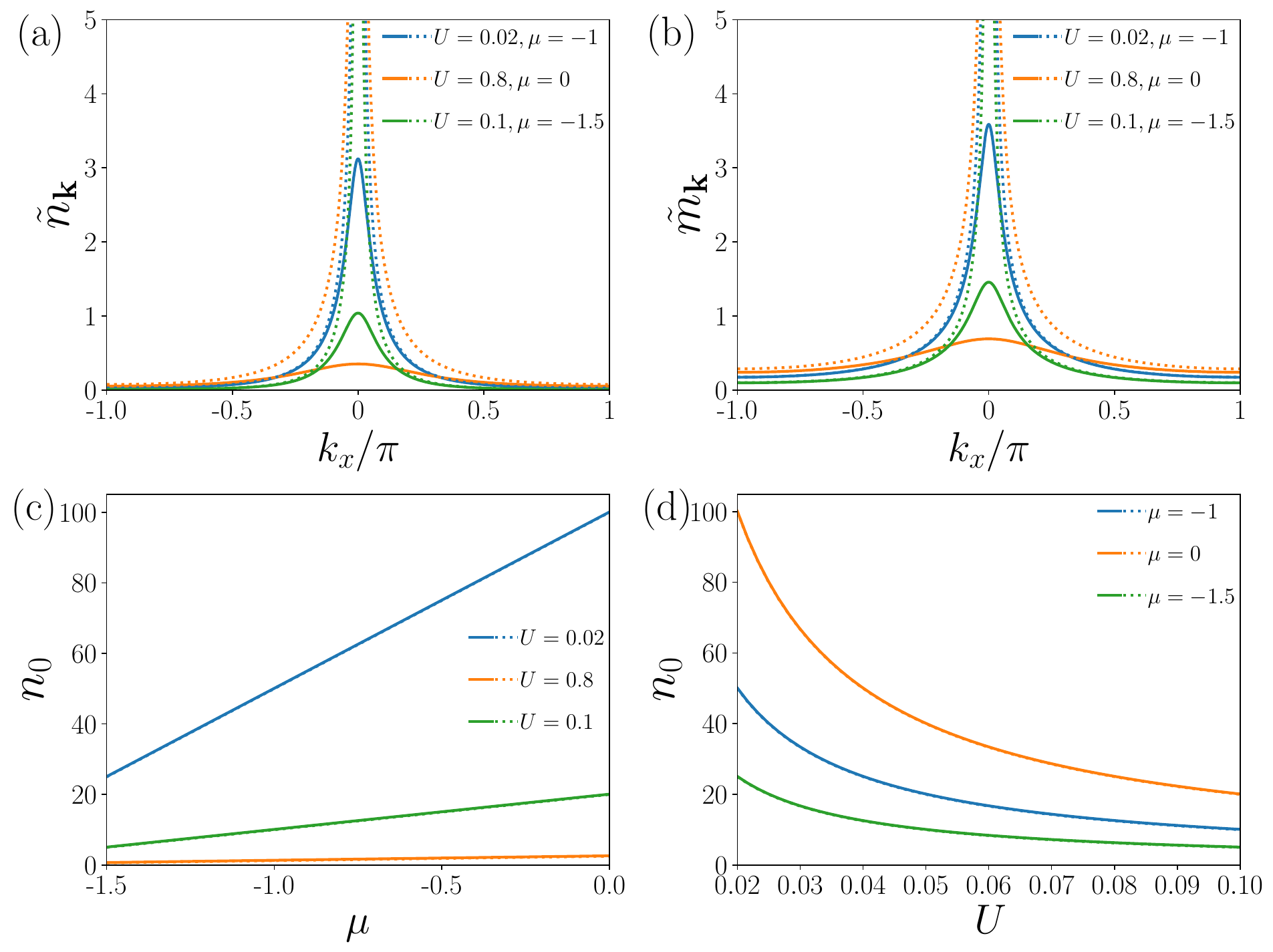}
    \caption{Comparison between the Gaussian variational principle and the Bogoliubov theory that describe the ground state of the Bose-Hubbard model deep in the superfluid phase. In all the panels, the solid and dotted lines correspond to the result of the Gaussian variational principle and the Bogoliubov theory, respectively. (a)~Mode occupation number of quantum depletions~\eqref{eq:n_k tilde} at $k_y=0$ as a function of $k_x$. (b)~Anomalous correlation function~\eqref{eq:m_k tilde} at $k_y=0$ as a function of $k_x$.
    (c)~Condensate fraction $n_0$ as a function of $\mu$ with fixed $U$. (d)~Condensate fraction $n_0$ as a function of $U$ with fixed $\mu$. 
    Note that, in (c) and (d), $n_0$ takes almost the same value for the Gaussian variational principle and for the Bogoliubov theory. 
    We set $L_y=50$ in all the plots.}
    \label{fig:n_k&m_k&n_0}
\end{figure}
\par 
In closing this subsection, we discuss the validity of the Bogoliubov theory. As we have seen, it is based on the assumption that the quantum depletions are dilute and the interaction between them is negligible because most of the bosons are in the condensate. 
Therefore, if we denote as $\tilde n_{\bf k}= \bra{\rm SF}b_{\bf k}^\dag b_{\bf k}\ket{\rm SF}$
the mode occupation number of quantum depletions, the validity of the Bogoliubov theory is guaranteed when $\sum_{\bf k}\tilde n_{\bf k}$ is sufficiently smaller than the total number of bosons $Vn_0+\sum_{\bf k}\tilde n_{\bf k}$. 
In other words, the Bogoliubov theory is valid when 
\begin{align}
    \frac{\sum_{\bf k}\tilde n_{\bf k}}{\sum_{\bf k}\tilde n_{\bf k}+Vn_0}
    \ll1
    \label{eq:gamma}. 
\end{align}
In the following, we restrict ourselves to the regime in which the left-hand side of the above inequality is smaller than $0.12$.

\subsection{Gaussian variational principle}

The Gaussian variational principle is a systematic extension of the Bogoliubov theory reviewed in the previous section. 
Although this approach is described in detail in Ref.~\cite{guaita-19}, where it is introduced for the first time, here we briefly review it for completeness. 
\par 
The Gaussian variational principle starts by assuming that the ground state deep in the superfluid phase $\ket{{\rm SF}}$ can be well described by a state belonging to the  manifold 
\begin{align}
    \mathcal{M}
    =
    \{
    \mathcal{D}(n_0)
    \mathcal{S}(\{\theta_{\bf k}\})\ket{0}
    \},
    \label{eq:Manifold GVP}
\end{align}
where $\mathcal{D}(n_0)$ and $\mathcal{S}(\{\theta_{\bf k}\})$ are the displacement and squeezing operators defined in Eqs.~\eqref{eq:displacement operator} and~\eqref{eq:squeezing operator}, respectively, and $\ket{0}$ is the vacuum of the bosonic operators $a_{{\bf i}}$.
We note that the manifold~\eqref{eq:Manifold GVP} contains the ground state predicted by the Bogoliubov theory~\eqref{eq:|SF> Bogoliubov}. 
The essential difference from the Bogoliubov theory is that, in the Gaussian variational principle, we determine the parameters $n_0$ and $\{\theta_{\bf k}\}$ imposing that the expectation value of the full Bose-Hubbard model, $\bra{\rm SF}H\ket{\rm SF}$, is minimum without neglecting the interaction between quantum depletions. 
As explicitly shown in Ref.~\cite{guaita-19}, this can be done by choosing these parameters as follows,
\begin{gather}
    n_0 
    = 
    \frac{2+\mu}{U}
    -2\tilde n -\tilde m, 
    \label{eq:n_0 GVP}
    \\
    \theta_{\bf k}
    =
    2{\rm sgn}(T_{\bf k})
    \tanh^{-1}
    \qty(
    \sqrt{\frac{1-\sqrt{1-T_{\bf k}^2}}{1+\sqrt{1-T_{\bf k}^2}}}
    ),
    \label{eq:Bogoliubov angle GVP}
\end{gather}
where 
\begin{gather}
    \tilde n = \frac{1}{V}\sum_{\bf k} \frac{\cosh\theta_{\bf k}-1}{2},
    \label{eq:tilde n}
    \\
    \tilde m = -\frac{1}{V}\sum_{\bf k} \frac{\sinh\theta_{\bf k}}{2},
    \label{eq:tilde m}
    \\
    T_{\bf k}
    = 
    \frac{\epsilon_{\bf 0}+2U\tilde n}
    {\epsilon_{\bf k}-2\epsilon_{\bf 0}-2U(\tilde n+\tilde m)}.
    \label{eq:T_k}
\end{gather}
By solving Eqs.~\eqref{eq:Bogoliubov angle GVP}-\eqref{eq:T_k} self-consistenly, we obtain the specific values of $n_0$ and $\{\theta_\mathbf{k}\}$ for a given $U$ and $\mu$. 
\par 
We note that, unlike the Bogoliubov theory, the Gaussian variational principle takes into account the interaction between quantum depletions because all the parameters are chosen so that the energy expectation value of the full Bose-Hubbard model is minimal. 
This fact appears as the correction by $\tilde n$ and $\tilde m$ in Eqs.~\eqref{eq:n_0 GVP} and~\eqref{eq:Bogoliubov angle GVP}. 
Indeed, if we take $\tilde n=\tilde m=0$ in these expressions, they reduce to the corresponding ones of the Bogoliubov theory in Eqs.~\eqref{eq:n_0} and~\eqref{eq:Bogoliubov angle}.  
Furthermore, thanks to this correction, the divergence of the Bogoliubov angle $\theta_{\bf k}$ in the long-wavelength limit ${\bf k}\to {\bf 0}$ is cured in the Gaussian variational principle. 
\par 
The validity of the Gaussian variational principle can be estimated using the fact that the manifold~\eqref{eq:Manifold GVP} includes the superfluid state predicted by the Bogoliubov theory~\eqref{eq:|SF> Bogoliubov} with Eqs.~\eqref{eq:n_0} and~\eqref{eq:Bogoliubov angle}.  
Since the superfluid state determined by the Gaussian variational principle is the lowest-energy state within the manifold~\eqref{eq:Manifold GVP}, it is expected to provide a better description than the one predicted by the Bogoliubov theory. 
Therefore, the Gaussian variational principle should be always valid whenever the Bogoliubov theory is also valid, i.e., whenever the condition~\eqref{eq:gamma} is satisfied. 
\par 
To see the difference between the Bogoliubov theory and the Gaussian variational principle more clearly, we plot in Figs.~\ref{fig:n_k&m_k&n_0}~(a) and~(b) the mode occupation number $\tilde n_{\bf k}$ and the anomalous correlation functions $\tilde m_{\bf k}$ of quantum depletions, which are respectively given by 
\begin{align}
    \tilde n_{\bf k}
    &=
    \bra{\rm SF}b_{\bf k}^\dag b_{\bf k}\ket{\rm SF}
    = 
    \frac{\cosh(\theta_{\bf k})-1}{2},
    \label{eq:n_k tilde}
    \\
    \tilde m_{\bf k}
    &= 
    \bra{\rm SF}b_{\bf k} b_{\bf -k}\ket{\rm SF}
    = 
    -
    \frac{\sinh(\theta_{\bf k})}{2}.
    \label{eq:m_k tilde}
\end{align}
These figures show that the Bogoliubov theory overestimates both $\tilde n_{\bf k}$ and $\tilde m_{\bf k}$ compared to the Gaussian variational principle. 
In particular, both quantities diverge in the long-wavelength limit ${\bf k}\to {\bf 0}$ within the Bogoliubov theory while they do not within the Gaussian variational principle. 
The overestimation can be understood as the result of neglecting the repulsive interaction between quantum depletions that suppresses their excitation. 
In Figs.~\ref{fig:n_k&m_k&n_0}~(c)~and~(d), we also plot the condensate fraction $n_0$ as a function of $U$ for several values of the chemical potential $\mu$ and vice-versa. We can see that the condensate fraction $n_0$ predicted by the Bogoliubov theory agrees well with that of the Gaussian variational principle. 

\section{Entanglement entropy}\label{sec:entanglement entropy}
In this section, we investigate the dynamics of the entanglement between the strip $A$ of Fig.~\ref{fig:torus} and the rest of the system after the quench from the superfluid phase of the Bose-Hubbard model~\eqref{eq:Bose-Hubbard Hamiltonian} to the free system~\eqref{eq:H_free}. We will take as initial superfluid the prediction both of the Bogoliubov theory and of the Gaussian variational principle reviewed in the previous section.

We employ as measure of entanglement the R\'enyi entanglement entropies, which are defined as 
\begin{align}
    S_n(t)
    = 
    \frac{1}{1-n}\log \Tr(\rho_A(t)^n), 
    \label{eq:RE}
\end{align}
where $n$ is the R\'enyi index. 
In the limit $n\to1$, Eq.~\eqref{eq:RE} gives the von Neumann entanglement entropy,
\begin{align}
    S_1(t)
    &=
    \lim_{n\to1}
    S_n(t)
    \nonumber\\
    &= 
    -\Tr(
    \rho_A(t)\log \rho_A(t)
    ).
\end{align}
The R\'enyi entanglement entropies in the superfluid phase of the 2D Bose-Hubbard model have been studied in Refs.~\cite{ahl-13, fr-16}.
\par
To calculate the time evolution of the R\'enyi entanglement entropies, we can exploit the fact that the initial superfluid state $\ket{\rm SF}$ is a coherent squeezed state both in the Bogoliubov theory and in the Gaussian variational principle. 
Since coherent squeezed states in general satisfy Wick's theorem and the post-quench Hamiltonian~\eqref{eq:H_free} is quadratic, the time-evolved reduced density matrix $\rho_A(t)$ is a bosonic Gaussian state, which satisfies Wick's theorem too. This implies that $\rho_A(t)$ is univocally characterized by its covariance matrix and mean vector~\cite{peschel-02, br-04, Banchi-2015}. 
To define the latter, it is convenient to introduce the canonical operators 
\begin{align}
    x_{\bf i}
    = 
    \frac{a_{\bf i}+a_{\bf i}^\dag}{\sqrt{2}},\quad 
    p_{\bf i}
    = 
    \frac{a_{\bf i}-a_{\bf i}^\dag}{\im \sqrt{2}}, 
\end{align}
with ${\bf i}\in A$. 
They satisfy the canonical commutation relation $[x_{\bf i},p_{\bf j}]=\im \delta_{\bf i,j}$. 
In terms of the vector $\boldsymbol{r}$ with entries $\boldsymbol{r}_{\bf i}=(x_{\bf i},p_{\bf i})^T$, these commutation relations can be cast in the compact form 
\begin{align}
    [\boldsymbol{r}_{\bf i},\boldsymbol{r}_{\bf i'}]=\im \Omega_{\bf i,i'},
\end{align}
where 
\begin{align}
    \Omega = \bigoplus_{i=1}^{V_A}\mqty(0&1\\-1&0), 
\end{align}
with $V_A=\ell L_y$ being the size of subsystem $A$. 
Then, $\rho_A(t)$ is univocally determined by the covariance matrix $\Gamma$ and the mean vector $\boldsymbol{s}$, whose entries are respectively defined as 
\begin{gather}
    \Gamma_{\bf i,i'}
    = 
    \frac{1}{2}
    \bra{\Psi(t)} \{{ \boldsymbol{r}_{\bf i}-\boldsymbol{s}_{\bf i},\boldsymbol{r}_{\bf i'}-\boldsymbol{s}_{\bf i'}}\} \ket{\Psi(t)},
    \\
    \boldsymbol{s}_{\bf i}
    = 
    \bra{\Psi(t)} \boldsymbol{r}_{\bf i}\ket{\Psi(t)}, 
\end{gather}
with ${\bf i,i'}\in A$.

If we consider the ground state $\ket{{\rm SF}}$ in the superfluid phase of the Bose-Hubbard model, determined either by the Bogoliubov theory or the Gaussian variational principle, and we quench to the free bosonic system~\eqref{eq:H_free}, the covariance matrix and the mean vector of the time-evolved state~\eqref{eq:time_ev} are of the form 
\begin{gather}
    \Gamma(t)_{\bf i,i'}
    = 
    \frac{1}{V}
    \sum_{\bf k}
    e^{-\im {\bf k\cdot(i-i')}}
    g_{\bf k}(t),
    \label{eq:CM}
    \\
    \boldsymbol{s}_{\bf i}
    = 
    \sqrt{2n_0}
    (1,0)^T, 
    \label{eq:MV}
\end{gather}
where 
\begin{align}
    g_{\bf k}(t)
    = 
    \frac{\cosh\theta_{\bf k}}{2}I
    + 
    \frac{\sinh\theta_{\bf k}}{2}
    \sigma_z e^{-2\im t \xi_{\bf k}\sigma_y},
    \label{eq:g_k(t)}
\end{align}
with $I$ and $\sigma_\nu$ ($\nu=x,y,z$) being the identity and the Pauli matrices, respectively, and $\xi_{\bf k}$ the single-particle dispersion relation of the free Hamiltonian~\eqref{eq:H_free}, determined in Eq.~\eqref{eq:H_free_diag}. The only difference in these expressions between choosing the Bogoliubov theory or the Gaussian variational principle is that in the former the angle $\theta_{\bf k}$ is given by Eq.~\eqref{eq:Bogoliubov angle} while in the latter is determined by Eq.~\eqref{eq:Bogoliubov angle GVP}.  
\par 
In terms of the covariance matrix $\Gamma(t)$, the R\'enyi entanglement entropies~\eqref{eq:RE} can be written as~\cite{cepd-05}
\begin{multline}
    S_n(t)
    = 
    \frac{1}{2(n-1)}
    \\
    \times
    \Tr\log\qty[\qty(\frac{\sqrt{W^2}+I}{2})^n-\qty(\frac{\sqrt{W^2}-I}{2})^n]. 
    \label{eq:RE exact}
\end{multline}
Here we have introduced the $2V_A\times 2V_A$ matrix $W(t)=-2\Gamma(t)\Omega$ whose entries are given by 
\begin{align}
    W(t)_{\bf i,i'}
    = 
    \frac{1}{V}
    \sum_{\bf k}
    e^{-\im {\bf k\cdot(i-i')}}
    w_{\bf k}(t),
    \label{eq:W_ii'}
\end{align}
where
\begin{align}
    w_{\bf k}(t)
    = 
    \cosh(\theta_{\bf k})\sigma_y 
    +
    \sinh(\theta_{\bf k})\sigma_z e^{-2\im t \xi_{\bf k}\sigma_y}.
    \label{eq:w_k}
\end{align}
Note that Eq.~\eqref{eq:RE exact} is exact regardless of the size and shape of the whole system and subsystem $A$. 
For a subsystem of finite size $V_A$, we can obtain the exact R\'enyi entanglement entropies by calculating numerically the eigenvalues of $W(t)$ and evaluating the trace in Eq.~\eqref{eq:RE exact} with them. 
We will check in this way the analytic results derived in the following.  
\par 
To obtain analytic expressions for the R\'enyi entanglement entropies after the quench, we consider as a subsystem $A$ the stripe of width $\ell$ of Fig.~\ref{fig:torus}. In this case, we can employ the dimensional reduction approach~\cite{cp-00}, previously applied to study entanglement at equilibrium in higher dimensional free fermionic and bosonic systems~\cite{aefs-14, mrc-20} and recently extended to analyze the entanglement entropies and the asymmetry after quenches in 2D free fermions~\cite{yac-24, yac-24-2}.
This method allows us to decompose the R\'enyi entanglement entropies of the two-dimensional system~\eqref{eq:RE exact} into the sum of the entropies of $L_y$ decoupled one-dimensional bosonic chains by exploiting the translational symmetry and the periodicity of the strip $A$ in the transverse direction. 
To this end, we introduce the unitary matrix $M$ with entries
\begin{align}
    M_{(i_x,i_y),(j_x,n_y)}
    = 
    \frac{\delta_{i_x,j_x}}{\sqrt{L_y}}
    e^{\im (2\pi n_y/L_y-\pi) i_y}. 
    \label{eq:M}
\end{align}
This matrix corresponds to performing a partial Fourier transform only in the transverse direction. Under it, the matrix~\eqref{eq:W_ii'} becomes block-diagonal,
\begin{align}
    MW(t)M^\dag 
    = 
    \bigoplus_{k_y}
    W_{k_y}(t), 
    \label{eq:W block diagonal}
\end{align}
where $W_{k_y}(t)$ is the $2\ell\times 2\ell$ block associated with the $k_y$ transverse momentum sector with entries
\begin{align}
    W_{k_y}(t)_{i_x,i_x'}
    =
    \frac{1}{L_x}
    \sum_{k_x}
    e^{-\im k_x(i_x-i_x')}
    w_{\bf k}(t). 
    \label{eq:W_ky}
\end{align}
Since the right-hand side of Eq.~\eqref{eq:RE exact} is invariant under unitary transformations on $W$, using the block diagonalization in Eq.~\eqref{eq:W block diagonal}, we can rewrite the R\'enyi entanglement entropies as the sum of the contribution of each transverse momentum sector,
\begin{align}
    S_{n}(t)
    = 
    \sum_{k_y} S_{n,k_y}(t),
\end{align}
where 
\begin{multline}
    S_{n,k_y}(t)
    = 
    \frac{1}{2(n-1)}
    \\
    \times
    \Tr\log\qty[\qty(\frac{\sqrt{W_{k_y}^2}+I}{2})^n-\qty(\frac{\sqrt{W_{k_y}^2}-I}{2})^n], 
    \label{eq:RE for k_y exact}
\end{multline}
are the R\'enyi entanglement entropies of an interval of length $\ell$ of a one-dimensional bosonic chain in a state described by the covariance matrix~\eqref{eq:W_ky}. 
\par 
Now the problem boils down to how to compute the one-dimensional R\'enyi entanglement entropies~\eqref{eq:RE for k_y exact}. 
To analytically calculate them, we first expand in Taylor series the right-hand side of Eq.~\eqref{eq:RE for k_y exact} in terms of $W_{k_y}$, 
\begin{gather}
    S_{n,k_y}(t)
    = 
    \frac{1}{2}
    \sum_{m=0}^\infty
    C_n(2m)
    \Tr(W_{k_y}(t)^{2m}),
    \label{eq:RE_exact_expand}
\end{gather}
where $C_n(2m)$ are the Taylor coefficients of the function 
\begin{align}
    h_n(x)
    = 
    \frac{1}{n-1}
    \log\left[
    \qty(
    \frac{|x|+1}{2}
    )^n
    -
    \qty(\frac{|x|-1}{2})^n
    \right].
\end{align}
As we will see in the following, the specific form of $C_n(2m)$ will never be needed and hence we do not report it.  
\par 
According to Eq.~\eqref{eq:W_ky}, $W_{k_y}(t)$ becomes a block-Toeplitz matrix generated by the symbol $w_{\bf k}(t)$ in the thermodynamic limit $L_x\to \infty$. 
In that case, the asymptotic form of the moments $\Tr(W_{k_y}(t)^{2m})$ in the ballistic regime, $t,\ell\to\infty$ with $\zeta=t/\ell$ fixed, can be analytically determined using the multidimensional stationary phase method introduced in Refs.~\cite{fc-08, cef-12} (see also Appendix~\ref{apdx:stationary_phase}) and reads 
\begin{multline}
    \Tr( W_{k_y}(t)^{2m})
    \simeq 
    2
    \int_{-\pi}^{\pi}
    \frac{{\rm d}k_x}{2\pi}
    \min(\ell,2|v_{k_x}|t) 
    \cosh^{2m}(\theta_{\bf k}).
    \label{eq:RE_stationary}
\end{multline}
Here $v_{k_x}=\partial_{k_x} \xi_{\bf k}$ is the longitudinal group velocity of the excitations in the free theory~\eqref{eq:H_free}. Inserting Eq.~\eqref{eq:RE_stationary} into~\eqref{eq:RE_exact_expand} and performing the sum over $k_y$, we eventually obtain 
\begin{align}
    S_{n}(t)
    \simeq 
    \sum_{k_y}
    \int_{-\pi}^\pi 
    \frac{{\rm d}k_x}{2\pi}
    \min(\ell,2|v_{k_x}|t)
    h_n(\cosh(\theta_{\bf k})).
    \label{eq:RE_QP}
\end{align}
According to Eq.~\eqref{eq:n_k tilde}, $\cosh(\theta_{\bf k})$ is related to the mode occupation of quantum depletions in the initial superfluid state by $\cosh(\theta_{\bf k})=2\tilde n_{\bf k}+1$. Observe that, in the 
large time limit, the entanglement entropy~\eqref{eq:RE_QP} tends to 
\begin{equation}
S_n(t\to\infty)\simeq \ell\sum_{k_y}\int_{-\pi}^{\pi} 
\frac{{\rm d}k_x}{2\pi}h_n(2\tilde{n}_{\bf k}+1),
\end{equation}
which coincides with the thermodynamic entropy of the generalized Gibbs ensemble determined by the mode occupations $\tilde{n}_{\bf k}$ of the quantum depletion in the initial superfluid state~\cite{ac-17-2},
\begin{equation}\label{eq:GGE}
\rho_{\rm GGE}=\frac{e^{-\sum_{\bf k}\lambda_{\bf k}b_{\bf k}^\dagger b_{\bf k}}}{\Tr(e^{-\sum_{\bf k}\lambda_{\bf k}b_{\bf k}^\dagger b_{\bf k}})},
\end{equation}
where $e^{\lambda_{\bf k}}=1+\tilde{n}_{\bf k}^{-1}$.

We can interpret Eq.~\eqref{eq:RE_QP} in terms of the quasiparticle picture for the quench dynamics of entanglement entropies~\cite{cc-05, ac-17, ac-17-2}. 
As Eq.~\eqref{eq:|SF> Bogoliubov} shows, quantum depletions with opposite momenta form condensed entangled pairs in the initial superfluid state. 
After the quench, each entangled pair is uncorrelated from the rest and propagates ballistically with velocity $\pm v_{k_x}$ in the longitudinal direction. Therefore, according to Eq.~\eqref{eq:RE_QP}, the entanglement entropy grows after the quench with the number of entangled pairs in the quantum depletion shared between $A$ and the rest of the system, which is given by the function $\min(2t|v_{k_x}|, \ell)$. Observe that the condensate part of the initial superfluid state does not play any role in the entanglement.
\par
\begin{figure}
    \raggedright
    \includegraphics[width=0.49\textwidth]{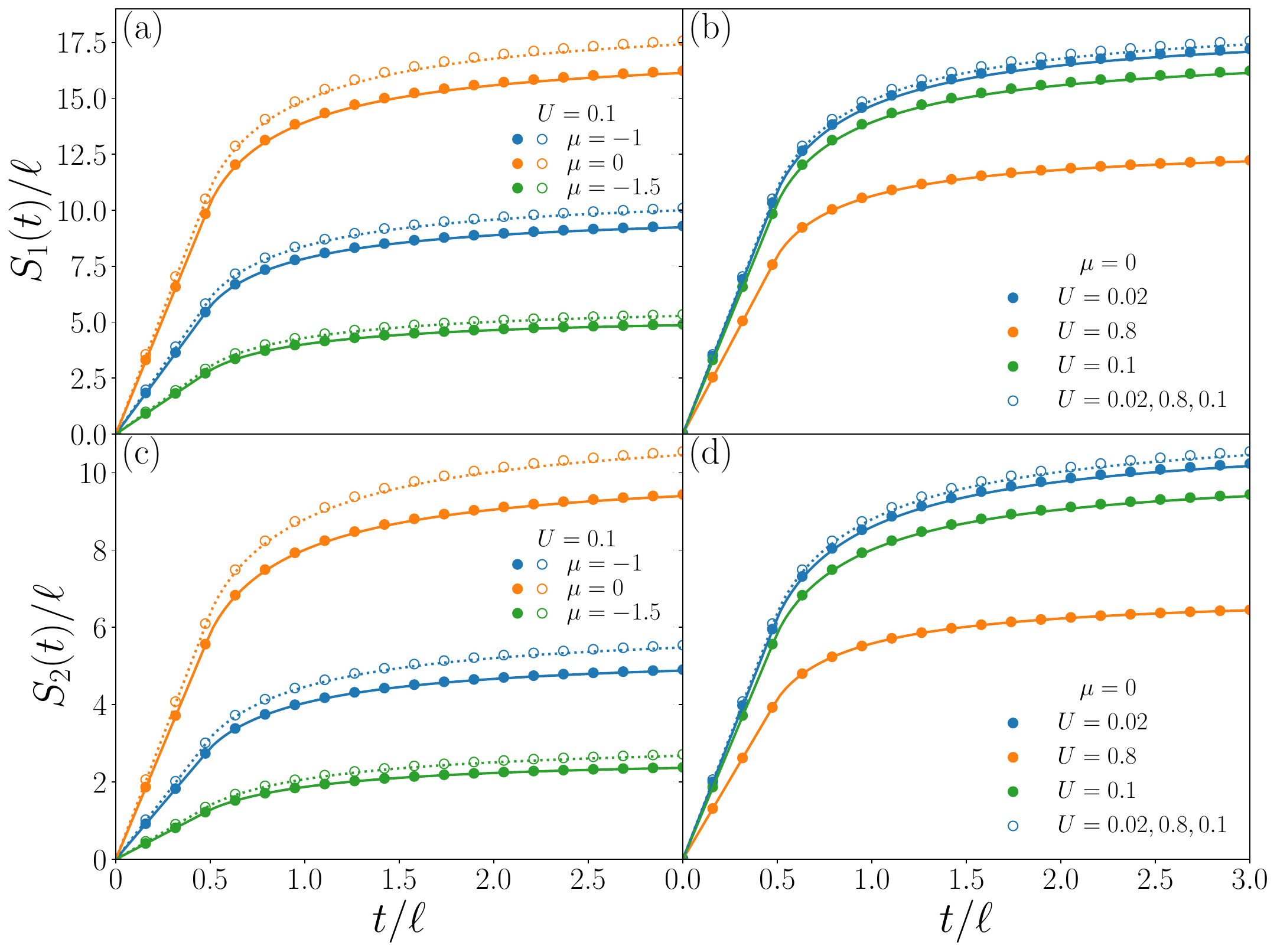}
    \caption{Time evolution of the von Neumann (upper panels) and $n=2$ R\'enyi (lower panels) entanglement entropies after a quantum quench from different points in the superfluid phase of the Bose-Hubbard model to the free-boson system. We take $L_x=2^{15}$, $L_y=50$, and $\ell=200$ in all the plots. The dashed and solid curves correspond to the analytic prediction~\eqref{eq:RE_QP} employing the Bogoliubov theory and the Gaussian variational principle for the superfluid initial state, respectively. The empty and filled circles are the exact value of the entanglement entropy for the Bogoliubov theory and Gaussian variational principle, respectively, obtained using Eq.~\eqref{eq:RE exact}.}
    \label{fig:entropies}
\end{figure}

 
In Fig.~\ref{fig:entropies}, we plot Eq.~\eqref{eq:RE_QP} as a function of time (curves) comparing it against the exact entanglement entropies (symbols) obtained numerically using Eq.~\eqref{eq:RE exact}. We take as initial superfluid state both the one predicted by the Bogoliubov theory (dashed curves and empty symbols) and the one determined by the Gaussian variational principle (solid curves and filled symbols). This figure shows that the analytic prediction~\eqref{eq:RE_QP} agrees well with the exact results. 
We find that in the quench from the Bogoliubov theory the entanglement entropies are in general larger compared to those obtained from the Gaussian variational principle. This effect is a consequence of the overestimation of the number of quantum depletions (see Fig.~\ref{fig:n_k&m_k&n_0}~(a)) due to neglecting the repulsive interaction between quantum depletions in the Bogoliubov theory.
As argued just before, the entangled pairs of quantum depletions are responsible for the entanglement growth after the quench and, consequently, the overestimation of them implies an overestimation of the post-quench entanglement entropies.
\par 
Figures~\ref{fig:entropies}~(b) and~(d) show that the R\'enyi entanglement entropies are independent of the initial on-site repulsive interaction $U$ in the Bogoliubov theory. The reason is again that in this case we are neglecting in the initial superfluid state the interaction between quantum depletions. Thus the mode occupation of quantum depletions in the initial state and, accordingly, the time-evolved R\'enyi entanglement entropies are independent of $U$. 

\section{Entanglement asymmetry}\label{sec:entanglement asymmetry}

As we already mentioned in Sec.~\ref{sec:effectiv theories}, the U(1) particle-number symmetry, which is generated by the conserved charge $Q=\sum_{\bf i}(a_{\bf i}^\dag a_{\bf i}+1/2)$, is spontaneously broken in the superfluid phase. 
In this section, we investigate this spontaneous symmetry breaking at the subsystem level and its fate after the quench to the free-boson Hamiltonian~\eqref{eq:H_free}, which respects this symmetry, using the entanglement asymmetry. Again we will compare the results obtained using the superfluid state predicted both by the Bogoliubov theory and the Gaussian variational principle. 

\subsection{Entanglement asymmetry for Gaussian states}
We focus on the R\'enyi  entanglement asymmetry of index $2$, defined as~\cite{amc-23}
\begin{align}
    \Delta S_A (t)
    = 
    \log[\Tr(\rho_A(t)^2)]
    -\log[\Tr(\rho_{A,Q}(t)^2)], 
    \label{eq:EA}
\end{align}
where $\rho_{A,Q}=\sum_{q} \Pi_q \rho_A \Pi_q$ is the symmetrization of $\rho_A$.
Here $\Pi_q$ is the projector onto the eigenspace of $Q_A$ with eigenvalue $q$, where $Q_A=\sum_{{\bf i}\in A}(a_{\bf i}^\dag a_{\bf i}+1/2)$ denotes the restriction of $Q$ to the subsystem $A$. This observable measures how much the symmetry generated by $Q$ is broken in the subsystem $A$. In fact, it is a positive definite quantity, $\Delta S_A(t)\geq 0$, being zero if and only if $[\rho_A,Q_A]=0$, i.e., when $\rho_A$ respects the U(1) particle-number symmetry~\cite{mhms-22, hms-23}. As we already mentioned in the introduction, the entanglement asymmetry has been recently used to study the quantum Mpemba effect via the dynamical restoration of the symmetry after the quench.
But it is also useful to examine the relaxation to non-Abelian generalized Gibbs ensembles when the symmetry is not restored~\cite{amvc-23} or confinement~\cite{khor-24}. It has been also extended to study 
the non-equilibrium properties of discrete symmetries~\cite{fac-24, cm-23} and non-internal symmetries such as spatial translations~\cite{krb-24}. The entanglement asymmetry of compact Lie groups has been further analyzed at equilibrium in matrix product states~\cite{cv-23}, in CFTs~\cite{chen-24, fadc-24, lmac-24}, and in random states~\cite{ampc-24}. 
\par 
To calculate the entanglement asymmetry~\eqref{eq:EA} in our setup, we consider the Fourier representation of the projector $\Pi_q$, 
\begin{align}
    \Pi_q 
    = 
    \int_{0}^{2\pi} 
    \frac{{\rm d}\alpha}{2\pi}
    e^{\im \alpha(Q_A-q)}.
\end{align}
If we plug it into Eq.~\eqref{eq:EA}, then we find 
\begin{align}
    \Delta S_A
    = 
    -
    \log 
    \int \limits_{[0,2\pi]^2}
    \frac{{\rm d}\alpha_1 {\rm d}\alpha_2}{(2\pi)^2}
    \frac{\Tr(\rho_A\rho_{A,\alpha_1-\alpha_2})}{\Tr(\rho_A^2)}, 
\end{align}
where $\rho_{A,\alpha}(t)=e^{\im \alpha Q_A} \rho_A(t) e^{-\im \alpha Q_A}$. 
Introducing the new variables $\chi_0=\alpha_1$ and $\chi = \alpha_2-\alpha_1$, it can be rewritten as 
\begin{align}
    \Delta S_A(t)
    = 
    -\log 
    \int_{R_\chi}
    \frac{{\rm d}\chi_0 {\rm d} \chi}{(2\pi)^2}
    \frac{Z_\chi(t)}{Z_0(0)}, 
    \label{eq:EA_2}
\end{align}
where 
\begin{align}
    Z_\chi(t)
    = 
    \Tr (\rho_A(t)\rho_{A,\chi}(t) )
    \label{eq:charged moment}
\end{align}
is usually dubbed charged moment. The integral domain $R_\chi$ in Eq.~\eqref{eq:EA_2} is defined as 
\begin{align}
    R_\chi:
    \chi_0\in[0,2\pi]\ {\rm and}\ \chi_0-\chi \in[0,2\pi].
\end{align}
Since the integrand of Eq.~\eqref{eq:EA_2} is independent of $\chi_0$, the integral over it can be easily calculated and we find
\begin{align}
    \Delta S_A(t) 
    = 
    -
    \log 
    \int_{-2\pi}^{2\pi}
    \frac{{\rm d}\chi}{2\pi}
    \qty(1-\frac{\abs{\chi}}{2\pi})
    \frac{Z_\chi (t)}{Z_0(t)}.
    \label{eq:EA_3}
\end{align}
If we take into account that the charged moments
$Z_\chi(t)$ are periodic, $Z_{\chi+2\pi}(t)=Z_\chi(t)$, then Eq.~\eqref{eq:EA_3} further simplifies,
\begin{equation}
\Delta S_A(t)=-\log\int_{-\pi}^{\pi}\frac{{\rm d}\chi}{2\pi}
\frac{Z_\chi(t)}{Z_0(t)}.
\end{equation}
\par 
We note that $\rho_{A,\chi}(t)$ is Gaussian because the particle-number operator $Q_A$ is quadratic. 
Therefore, as happens for $\rho_A(t)$, $\rho_{A,\chi}(t)$ can also be univocally characterized by its covariance matrix $\Gamma_{\chi}(t)$ and mean vector $\boldsymbol{s}_{\chi}$. Their entries are related to those of the covariance matrix and mean vector of $\rho_A(t)$, determined in Eqs.~\eqref{eq:MV} and~\eqref{eq:g_k(t)}, by 
\begin{gather}
    \Gamma_\chi(t)_{\bf i,i'}
    =
    e^{-\im \chi \sigma_y}\Gamma(t)_{\bf i,i'}e^{\im \chi \sigma_y},     \label{eq:Gamma_chi}
    \\
    (\boldsymbol{s}_\chi)_{\bf i}
    = 
    e^{-\im\chi\sigma_y}\boldsymbol{s}_{\bf i}. 
    \label{eq:s_chi}
\end{gather}
\par 
Since both $\rho_A(t)$ and $\rho_{A,\chi}(t)$ are Gaussian, the charged moments~\eqref{eq:charged moment} are the overlap of two bosonic Gaussian states. 
As explicitly shown in Appendix~\ref{apdx:charged_moment}, the Wigner function formalism~\cite{bvl-05} allows us to rewrite the overlap of these Gaussian states in terms of their covariance matrices and mean vectors as 
\begin{align}
    Z_\chi(t)
    = 
    \frac{\exp[-\frac{1}{2}(\boldsymbol{s}-\boldsymbol{s}_\chi)^T(\Gamma+\Gamma_\chi)^{-1}(\boldsymbol{s}-\boldsymbol{s}_\chi)]}{\sqrt{\det(\Gamma+\Gamma_\chi)}}. 
    \label{eq:charged moment exact}
\end{align}
\par 
This expression is exact and valid for any geometry and size of the subsystem $A$. When $\chi=0$, we recover the 
formula for the neutral moments $\Tr(\rho_A^2)$ obtained
in Eq.~\eqref{eq:RE exact}.

As we have done in Sec.~\ref{sec:entanglement entropy} for the R\'enyi entanglement entropies, from now on we take as subsystem $A$ the strip of width $\ell$ in Fig.~\ref{fig:torus} and we apply the dimensional reduction approach to the charged moment~\eqref{eq:charged moment exact} using the partial Fourier transform~\eqref{eq:M}. 
According to Eq.~\eqref{eq:charged moment exact}, $Z_{\chi}(t)$ is invariant under this unitary transformation, $(\boldsymbol{s}-\boldsymbol{s}_\chi)\mapsto M(\boldsymbol{s}-\boldsymbol{s}_\chi)$ and $(\Gamma+\Gamma_\chi)\mapsto M(\Gamma+\Gamma_\chi)M^\dag$. 
Combining Eqs.~\eqref{eq:M}, \eqref{eq:Gamma_chi}, and \eqref{eq:s_chi}, we specifically obtain 
\begin{equation}\label{Gamma+Gamma}
M(\Gamma(t)+\Gamma_{\chi}(t))M^\dagger=\bigoplus_{k_y}e^{-\frac{\chi}{2}\Omega}\Lambda_{k_y, \chi}(t)e^{\frac{\chi}{2}\Omega},
\end{equation}
where $\Lambda_{k_y,\chi}(t)$ is the $2\ell \times 2\ell$ matrix with entries
\begin{align}
    \Lambda_{k_y,\chi}(t)_{i_x,i_x'}
    = 
    \frac{1}{L_x}
    \sum_{k_x}
    e^{-\im k_x(i_x-i_x')}
    \lambda_{\bf k}(t), 
\end{align}
and symbol 
\begin{align}
    \lambda_{\bf k}(t)
    = 
    \cosh(\theta_{\bf k}) I 
    + 
    \sinh(\theta_{\bf k})\cos(\chi)
    \sigma_z e^{-2\im t \xi_{\bf k}\sigma_y}.
    \label{eq:lambda_k}
\end{align}
The entries of the transformed mean vector are
\begin{equation}
[M(\boldsymbol{s}-\boldsymbol{s}_\chi)]_{(i_x,n_y)}
    = 
    \delta_{k_y,0}
    \sqrt{8n_0L_y}\sin(\frac{\chi}{2})
    e^{-\frac{\im}{2}\chi \sigma_y}
    \boldsymbol{u}_{i_x},
    \label{eq:s-s}
\end{equation}
where $\boldsymbol{u}_{i_x}=(0, 1)^T$.
Applying Eqs.~\eqref{eq:s-s} and~\eqref{Gamma+Gamma} in Eq.~\eqref{eq:charged moment exact}, we find
\begin{align}
    Z_\chi(t) 
    =
    \frac{
    \exp(-4n_0L_y\sin^2(\frac{\chi}{2}) 
    \boldsymbol{u}^T
    [\Lambda_{0,\chi}(t)]^{-1}
    \boldsymbol{u})
    }
    {
    \prod_{k_y}\sqrt{\det(\Lambda_{k_y,\chi}(t))}
    },
    \label{eq:charged moment dimensional reduction}
\end{align}
where $\boldsymbol{u}$ is $2\ell$-dimensional vector whose elements are $\boldsymbol{u}=(\boldsymbol{u}_{1}, \dots, \boldsymbol{u}_{\ell})^T$.
Plugging Eq.~\eqref{eq:charged moment dimensional reduction} into Eq.~\eqref{eq:EA_3}, we arrive at 
\begin{align}
    \Delta S_A(t)
    =
    -\log 
    \int_{-\pi}^{\pi}
    \frac{{\rm d}\chi}{2\pi}
    e^{-V_A[A_\chi(t)+B_\chi(t)]}, 
    \label{eq:EA_exact_final}
\end{align}
with 
\begin{align}
    A_\chi(t)
    &=
    \frac{1}{2V_A}
    \sum_{k_y}
    \log(\frac{\det[\Lambda_{k_y,\chi}(t)]}{\det[\Lambda_{k_y,0}(t)]}),
    \label{eq:A}
    \\
    B_\chi(t)
    &=
    \frac{4n_0\sin^2(\frac{\chi}{2})}{\ell}\boldsymbol{u}^T 
    [\Lambda_{0,\chi}(t)]^{-1}\boldsymbol{u}.
    \label{eq:B}
\end{align}
Equation~\eqref{eq:EA_exact_final} is exact for any size of the strip $A$ and we will thus employ it to numerically benchmark the analytic results that we find in the rest of this section.

\subsection{Entanglement asymmetry of the superfluid state}
Exploiting the previous machinery, let us first study the entanglement asymmetry of the ground state deep in the superfluid phase of the Bose-Hubbard model, which is also the initial state in the quench protocol. 

As we have just seen, the entanglement asymmetry in the strip $A$ can be exactly calculated from Eq.~\eqref{eq:EA_exact_final}. In what follows, we will derive using this expression its asymptotic behavior for $\ell\gg1$ taking the thermodynamic limit in the longitudinal direction $L_x\to \infty$. 
 According to Eq.~\eqref{eq:A}, the term $A_\chi(0)$  is given by the determinant of $\Lambda_{k_y,\chi}(0)$, which becomes a block-Toeplitz matrix generated by the symbol~\eqref{eq:lambda_k} in that limit. 
The leading term of the determinant of a block-Toeplitz matrix for $\ell\gg1$ can be calculated using the Widom-Szeg\"o theorem~\cite{widom}, 
\begin{align}\label{eq:A(0) Wisdom-Szego}
    \log\det(\Lambda_{k_y,\chi})\simeq 
    \ell 
    \int_{-\pi}^\pi 
    \frac{{\rm d}k_x}{2\pi}
    \log\det(\lambda_{\bf k}(0)).
\end{align}
Employing this result in Eq.~\eqref{eq:A}, we obtain 
\begin{multline}
    A_\chi(0)
    \simeq 
    \frac{1}{L_y}
    \sum_{n_y=0}^{L_y-1}
    \int_{-\pi}^\pi
    \frac{{\rm d}k_x}{4\pi}
    \\
    \times
    \log(\cosh^2(\theta_{\bf k})-\sinh^2(\theta_{\bf k}) \cos^2(\chi)).
    \label{eq:A(0)}
\end{multline}
\par 
To calculate the term $B_\chi(0)$, which is given by Eq.~\eqref{eq:B}, we have to evaluate the inverse of the matrix $\Lambda_{0,\chi}(0)$. 
Although the inverse of a block-Toeplitz matrix is in general not a block-Toeplitz, it can be approximated for $\ell\gg1$  by the block-Toeplitz matrix generated by the inverse of the symbol~\cite{amvc-23}, i.e,  
\begin{align}\label{eq:inverse}
    [\Lambda_{0,\chi}(0)]^{-1}_{i_x,i_x'}
    \simeq 
    \int_{-\pi}^\pi
    \frac{{\rm d}k_x}{2\pi}
    e^{-\im k_x(i_x-i_x')}
    [\lambda_{(k_x,0)}(0)]^{-1}.
\end{align}
Applying this approximation in Eq.~\eqref{eq:B}, we find 
\begin{align}
    B_\chi(0)
    \simeq 
    \int_{-\pi}^\pi
    {\rm d}k_{x}
    \frac{4n_0 \sin^2(\chi/2)d_\ell(k_{x})}{\cosh(\theta_{k_x})-\sinh (\theta_{k_x})\cos \chi},  
    \label{eq:B(0)}
\end{align}
where $\theta_{k_x}\equiv\theta_{(k_x,0)}$ and 
\begin{align}
    d_\ell(k)
    = 
    \frac{\sin^2(k\ell/2)}{2\pi \ell \sin^2(k/2)}.
\end{align}
The latter tends to the Dirac delta function in the limit $\ell\to\infty$, i.e. $\lim_{\ell \to \infty}d_{\ell}(k)=\delta(k)$. 
Note that Eqs.~\eqref{eq:A(0)} and~\eqref{eq:B(0)} are valid for the state predicted by the Gaussian variational principle but they are ill-defined in the Bogoliubov theory since $\lim_{k_x\to0}\theta_{k_x}=-\infty$ in the thermodynamic limit $L_x\to\infty$. 
\par 
Inserting the above results for $A_\chi(0)$ and $B_\chi(0)$ into Eq.~\eqref{eq:EA_exact_final} and integrating over $\chi$, we can deduce the asymptotic behavior of the entanglement asymmetry. 
Since the exponent of the integrand in Eq.~\eqref{eq:EA_exact_final} is proportional to the subsystem size $V_A$, this integral can be evaluated when $V_A\gg1$ using the saddle point approximation. The saddle points of the integrand correspond to the solutions of the equation 
\begin{align}
    \partial_\chi[A_\chi(0)+B_\chi(0)]=0. 
\end{align}
By solving it employing Eqs.~\eqref{eq:A(0)} and~\eqref{eq:B(0)}, one finds that, in the integration domain, there is a saddle point at $\chi=0$. 
Therefore, expanding the exponent of the integrand in Eq.~\eqref{eq:EA_exact_final} up to the second order in $\chi$ around this point, we can approximate it as 
\begin{align}
    \Delta S_A(0)
    \simeq 
    -\log 
    \int_{-\infty}^\infty
    \frac{{\rm d}\chi}{2\pi}
    e^{-V_A [\alpha(0)+\beta(0)] \chi^2}, 
    \label{eq:EA_t=0}
\end{align}
with 
\begin{gather}
    \alpha(0)
    = 
    \frac{1}{2}
    \partial_\chi^2 A_\chi(0)|_{\chi=0}
    \simeq 
    \frac{2}{L_y}
    \sum_{k_y}
    \int_{-\pi}^\pi
    \frac{{\rm d}k_x}{2\pi}
    \abs{\tilde m_{\bf k}}^2,
    \label{eq:alpha_0}
    \\
    \beta(0)
    =
    \frac{1}{2}\partial_\chi^2 B_\chi(0)|_{\chi=0}
    \simeq 
    n_{0}
    \int_{-\pi}^\pi
    {\rm d}k_x
    d_\ell(k_x) e^{-|\theta_{k_x}|}.
    \label{eq:beta_0}
\end{gather}
The Gaussian integral in Eq.~\eqref{eq:EA_t=0} can be calculated by standard means and we finally arrive at the compact expression 
\begin{align}
    \Delta S_A(0)
    \simeq 
    \frac{1}{2}
    \log[4\pi V_A (\alpha(0)+\beta(0))].
    \label{eq:EA_SF}
\end{align}
Again, note that this result is only valid for the Gaussian variational principle. 
\par 
The analytic prediction~\eqref{eq:EA_SF} shows that the entanglement asymmetry grows with the logarithm of the 
size of the strip $A$. The same behavior has been found 
for the ground state of free fermions on a 2D lattice~\cite{yac-24}. The term independent of the subsystem size consists of two contributions $\alpha(0)$ and $\beta(0)$. Since $\tilde m_{\bf k}$, defined in Eq.~\eqref{eq:alpha_0}, is the anomalous correlation function between quantum depletions with opposite momenta, $\alpha(0)$ encodes the contribution of the quantum depletions in the superfluid state. 
On the other hand, $\beta(0)$ in Eq.~\eqref{eq:beta_0} is proportional to the condensate fraction $n_0$ and, therefore, corresponds to the contribution of the Bose-Einstein condensate. 
Observe that $\beta(0)$ is suppressed exponentially by the factor $e^{-\abs{\theta_{k_x}}}$, which may be interpreted as the result of the interaction between quantum depletions and the condensate. 
Since $d_\ell(k_x)$ tends to a Dirac delta at $k_x=0$ when $\ell\to\infty$, only quantum depletions with small momentum are relevant to this suppression. 
\begin{figure}
    \raggedright
    \includegraphics[width=0.45\textwidth]{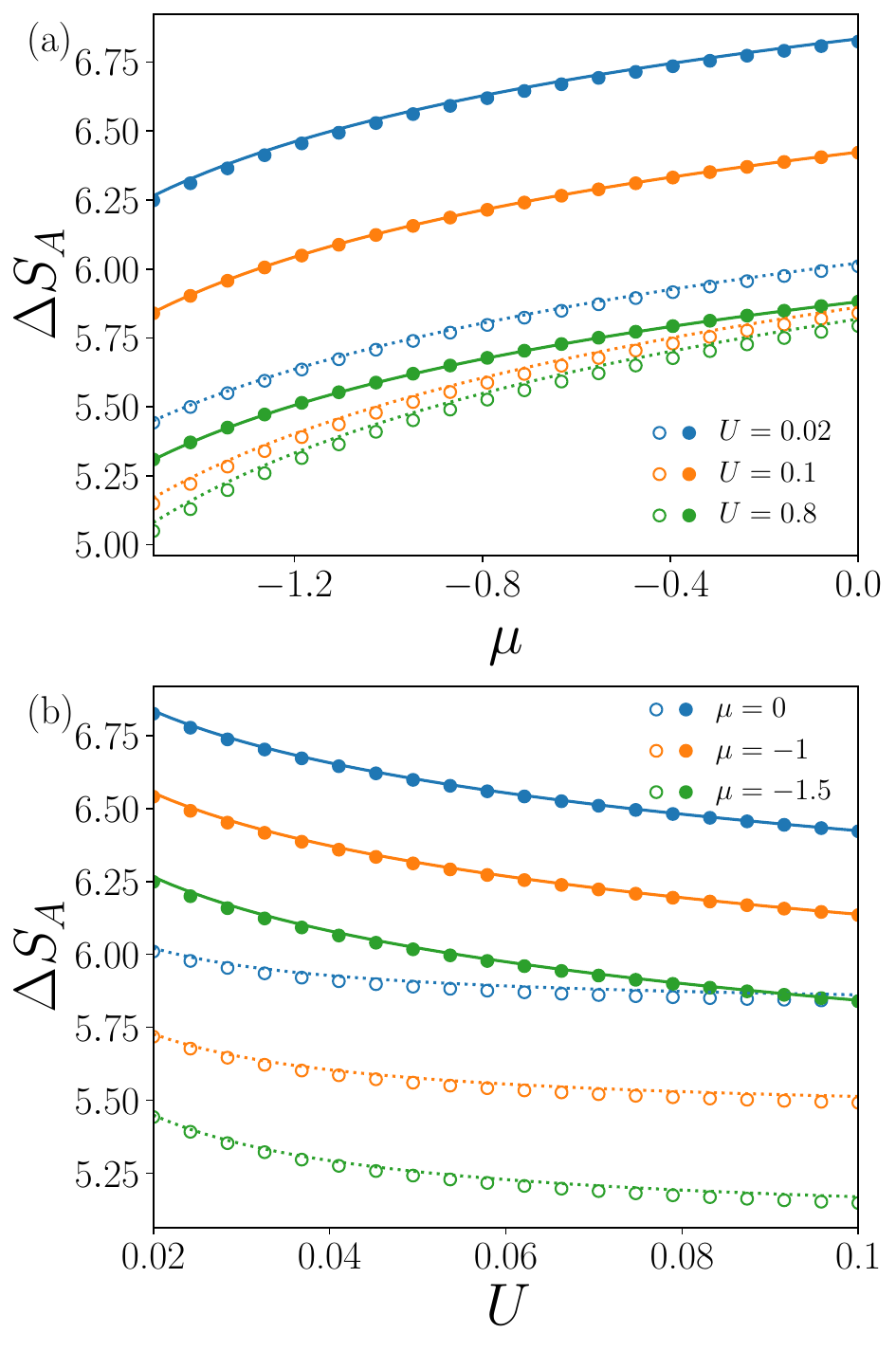}
    \caption{Entanglement asymmetry for the ground state of the Bose-Hubbard model deep in the superfluid phase as a function of $\mu$ and several fixed values of $U$ (upper panel) and vice-versa (lower panel). 
    The empty and filled circles are the exact asymmetry obtained numerically using Eq.\,\eqref{eq:EA_exact_final} for the Bogoliubov theory and the Gaussian variational principle, respectively. 
    The solid and dashed lines correspond to the analytic prediction \eqref{eq:EA_SF}. 
    For the Gaussian variational principle, we employ Eqs.~\eqref{eq:alpha_0} and \eqref{eq:beta_0} to determine $\alpha(0)$ and $\beta(0)$, while for the Bogoliubov theory, we evaluate them calculating numerically $A_\chi(0)$ and $B_\chi(0)$, respectively. 
    We set $\ell=200$ and $L_y=50$ in all the plots. }
    \label{fig:EA_SF}
\end{figure}
\par
In Fig.~\ref{fig:EA_SF}, we plot the entanglement asymmetry of the superfluid state as a function of the repulsive interaction $U$ for several fixed values of the chemical potential $\mu$ (upper panel) and vice-versa (bottom panel). We compare the analytic prediction~\eqref{eq:EA_SF} for the Gaussian variational state (solid lines) with the exact values obtained numerically from Eq.~\eqref{eq:EA_exact_final} (filled symbols).  
In this case, the analytic expressions agree well with the exact results. The empty symbols correspond to the exact value of the entanglement asymmetry in the Bogoliubov theory calculated with Eq.~\eqref{eq:EA_exact_final}. As we already remarked the asymptotic expression~\eqref{eq:EA_SF} cannot be applied in this case using Eqs.~\eqref{eq:alpha_0} and~\eqref{eq:beta_0} for $\alpha(0)$ and $\beta(0)$. Instead, the dashed lines correspond to Eq.~\eqref{eq:EA_SF} but determining  $\alpha(0)$ and $\beta(0)$ numerically from Eqs.~\eqref{eq:A} and~\eqref{eq:B}.

As expected, Fig.~\ref{fig:EA_SF} shows that the entanglement asymmetry increases as one goes deeper into the superfluid phase, i.e., as $U$ decreases or $\mu$ increases. 
This behavior is similar to the superfluid order parameter $n_0$ shown in Figs.~\ref{fig:n_k&m_k&n_0} (c) and (d). 
However, the entanglement asymmetry contains more information for characterizing the symmetry breaking compared to the local order parameter in the sense that, as shown by Eq.~\eqref{eq:EA_SF}, it includes not only the contribution of the condensate but also of the quantum depletions, which are also responsible for breaking the particle-number symmetry. 

\subsection{Time evolution of the entanglement asymmetry after the quench}

Let us now investigate the time evolution of the entanglement asymmetry when we perform a sudden quench from the superfluid state to the free-boson Hamiltonian~\eqref{eq:H_free}. 
\par
We can derive analytic expressions for the entanglement asymmetry in the ballistic limit $t,\ell\to\infty$ with $\zeta=t/\ell$ fixed from Eq.~\eqref{eq:EA_exact_final} using the multidimensional stationary phase approximation~\cite{fc-08, cef-12}. The calculations are quite cumbersome and are detailed in Appendix~\ref{apdx:mult_st_app} and~\ref{apdx:B_chi}, here we report the final results. For the 
term $A_\chi(t)$ in Eq.~\eqref{eq:EA_exact_final}, we obtain (see Appendix~\ref{apdx:mult_st_app})

\begin{multline}
    A_\chi(t)
    \simeq 
    \frac{1}{L_y}
    \sum_{k_y}
    \int_{-\pi}^\pi 
    \frac{{\rm d}k_x}{4\pi}
    x_{k_x}(\zeta)
    \\
    \times 
    \log(\cosh^2(\theta_{\bf k})-\sinh^2(\theta_{\bf k})\cos^2(\chi)),
    \label{eq:A(t) analytic}
\end{multline}
where $x_{k_x}(\zeta)=1-\min(2\zeta |v_{k_x}|,1)$. For the term $B_\chi(t)$, we find (see Appendix~\ref{apdx:B_chi})

\par 

\begin{multline}
    B_\chi(t)\simeq 
    4n_0 \sin^2\left(\frac{\chi}{2}\right)
    \int_{-\pi}^\pi 
    {\rm d}k_x d_\ell(k_x)
    \\
    \times 
    \Bigg[
    \frac{x_{k_x}(\zeta)[\cosh(\theta_{k_x})+\sinh(\theta_{k_x})\cos(\chi)\cos(2t\xi_{k_x})]}{\cos^2(\theta_{k_x})-\sinh^2(\theta_{k_x})\cos^2(\chi)}
    \\
    +
    \frac{1-x_{k_x}(\zeta)}{\cosh\theta_{k_x}}
    +
    \frac{[1-x_{k_x}(\zeta)]\sinh(\theta_{k_x})\cos(\chi)\cos(2t\xi_{k_x})}{\cosh^2\theta_{k_x}}
    \Bigg].
    \label{eq:B(t) analytic}
\end{multline}
\par 
The entanglement asymmetry can be now determined by inserting Eqs.~\eqref{eq:A(t) analytic} and~\eqref{eq:B(t) analytic} into Eq.~\eqref{eq:EA_exact_final} and integrating over $\chi$. 
As we have done when we derived the entanglement asymmetry at $t=0$ in Eq.~\eqref{eq:EA_SF}, we can evaluate the integral over $\chi$ using the saddle point approximation. Following the same procedure, we obtain that the entanglement asymmetry at finite $t$ is 
\begin{align}
    \Delta S_A(t)
    \simeq 
    \frac{1}{2}
    \log[4\pi V_A(\alpha(t)+\beta(t))], 
    \label{eq:EA(t)}
\end{align}
where 
\begin{align}
    \alpha(t)
    &= 
    \frac{1}{2}\partial_\chi^2 A_\chi(t)|_{\chi=0}
    \label{eq:alpha_t_numerics}
    \\
    &\simeq 
    \frac{2}{L_y}
    \sum_{k_y}
    \int_{-\pi}^\pi
    \frac{{\rm d}k_x}{2\pi}
    x_{k_x}(\zeta)
    \abs{m_{\bf k}}^2
    \label{eq:alpha_t}
\end{align}
and
\begin{align}
    \beta(t)
    &=
    \frac{1}{2}\partial_\chi^2 B_\chi(t)|_{\chi=0} \label{eq:beta_t_numerics}
    \\
    &\simeq 
    n_0
    \int_{-\pi}^\pi
    {\rm d}k_x
    d_\ell (k_x)
    \Bigg[
    x_{k_x}(\zeta)+\frac{1-x_{k_x}(\zeta)}{\cosh(\theta_{k_x})^2}\Bigg]
    \nonumber\\
    &\quad \times 
    [\cosh(\theta_{k_x})+\sinh(\theta_{k_x})\cos(2t\xi_{k_x})]
    .\label{eq:beta_t}
\end{align}
If we set $t=0$, then Eq.~\eqref{eq:EA(t)} reduces to Eq.~\eqref{eq:EA_SF}. 
\par 
According to Eq.~\eqref{eq:EA(t)}, the entanglement asymmetry after the quench is given by the time-dependent functions $\alpha(t)$ and $\beta(t)$. 
The former can be understood in terms of the quasiparticle picture for the quantum depletions as follows.
As already discussed in Sec.~\ref{sec:entanglement entropy}, pairs of entangled quantum depletions in the initial state propagate ballistically in opposite directions with longitudinal velocity $\pm v_{k_x}$ after the quench. Since the anomalous correlation $\tilde{m}_{\bf k}$ is not zero, these pairs break the particle-number symmetry. After the quench, they contribute to the entanglement asymmetry in $A$ as long as both excitations are inside the subsystem. At the instant in which one of the two excitations leaves $A$, the pair does not contribute anymore to breaking the symmetry in the subsystem, since the function $x_{k_x}(\zeta)$ in Eq.~\eqref{eq:alpha_t} counts the number of pairs with both excitations inside the subsystem. 
It is easy to check that, as time passes, the number of complete pairs in $A$ decreases and is zero in the limit $t\to\infty$. This implies that their contribution in the stationary states vanishes, i.e. $\alpha(t\to\infty)=0$. 
\par 
This interpretation for $\alpha(t)$ is quite similar to the quasiparticle picture for the quench dynamics of entanglement asymmetry in free fermion systems, both in 1D~\cite{makc-24, carc-24, rylands-24} and 2D~\cite{yac-24-2}. In that case, the particle number symmetry is broken due to the presence of Cooper pairs in the initial state. After the quench, these pairs propagate ballistically and only those with both excitations in the subsystem count for the entanglement asymmetry, as happens here for the quantum depletions. In general,
the number of complete Cooper pairs inside a subsystem reduces to zero at large times and the symmetry is restored. There are more exotic situations in which the initial state contains multiplets, instead of pairs, and the subsystem tends to a non-Abelian generalized Gibbs ensemble, spoiling the restoration of the symmetry~\cite{amvc-23, cma-24}.  
\par 
The peculiar point in our present case is that the entanglement asymmetry~\eqref{eq:EA(t)} contains the additional contribution $\beta(t)$ proportional to the condensate fraction $n_0$, which cannot be explained in terms of the quasiparticle picture. 
In particular, this term does not vanish but converges to the finite value at large times 
\begin{align}
    \beta(\infty)
    =
    n_0 \int_{-\pi}^\pi {\rm d}k_x
    \frac{d_\ell(k_x)}{\cosh(\theta_{k_x})}.
    \label{eq:beta(inf)}
\end{align}
This means that the entanglement asymmetry also tends at large times to the non-zero value 
\begin{align}
    \Delta S_A(\infty)
    \simeq 
    \frac{1}{2}
    \log(4\pi V_A\beta(\infty)). 
\end{align}
Therefore, the U(1) particle-number symmetry spontaneously broken in the initial superfluid state is not restored in the subsystem after the quench, even though the post-quench Hamiltonian~\eqref{eq:H_free} does respect the symmetry. 
From Eq.~\eqref{eq:beta(inf)}, we can conclude that the lack of symmetry restoration is clearly due to the condensate, whose fraction remains non-zero after the quench. 
\par
\begin{figure}
    \raggedright
    \includegraphics[width=0.45\textwidth]{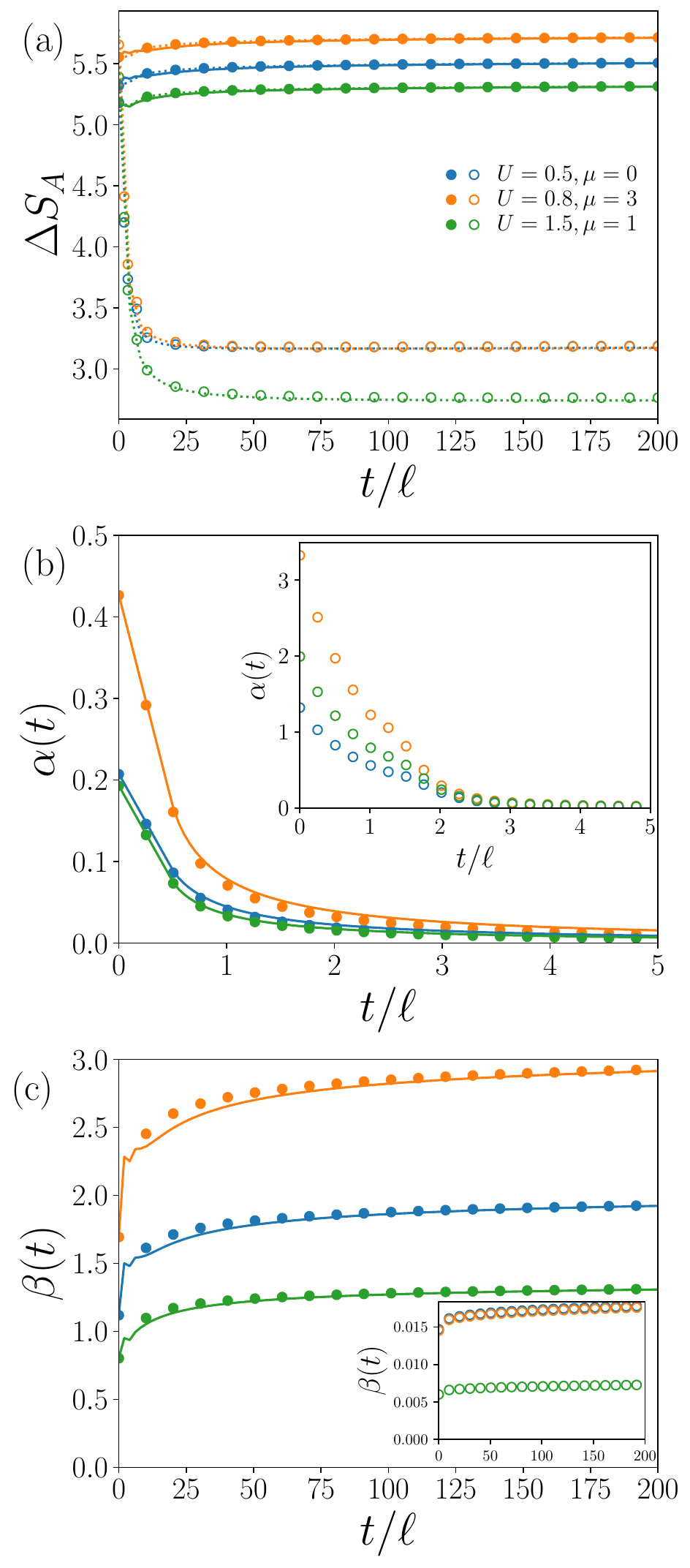}
    \caption{(a) Time evolution of the entanglement asymmetry after a quench to the free bosonic system from different superfluid states. The empty and filled circles are the exact values obtained using Eq.~\eqref{eq:EA_exact_final} with the Bogoliubov theory and the Gaussian variational principle, respectively. 
    The solid lines are the analytic prediction \eqref{eq:EA(t)} with Eqs.~\eqref{eq:alpha_t} and \eqref{eq:beta_t} for the Gaussian variational principle. 
    The dotted lines correspond to Eq.~\eqref{eq:EA(t)} with $\alpha(t)$ and $\beta(t)$ exactly calculated using Eqs.~\eqref{eq:alpha_t_numerics} and \eqref{eq:beta_t_numerics}. 
    (b) and (c) Time evolution of the contributions to the asymmetry from the quantum depletions, $\alpha(t)$, and from the condensate, $\beta(t)$. 
    The empty and filled circles are the exact results for the Bogoliubov theory and the Gaussian variational principle, respectively. The solid lines are the analytic predictions~\eqref{eq:alpha_t} and \eqref{eq:beta_t}. 
    We set $L_y=\ell=50$ in all the plots.     }
    \label{fig:EA}
\end{figure}

 
In Fig.~\ref{fig:EA}~(a), we represent the entanglement asymmetry after the quench as a function of time for different initial superfluid states. The symbols represent the exact value of the asymmetry calculated with Eq.~\eqref{eq:EA_exact_final} taking the initial state of the Bogoliubov theory (empty circles) and of the Gaussian variational principle (filled circles). We plot the analytic prediction~\eqref{eq:EA(t)} with Eqs.~\eqref{eq:alpha_t} and~\eqref{eq:beta_t} only for the Gaussian variational principle (solid lines). 
In the Bogoliubov theory, these expressions are problematic because, in the thermodynamic limit $L_x\to\infty$,  $\theta_{k_x}$ diverges. Instead, in that case we plot Eq.~\eqref{eq:EA(t)} with the exact $\alpha(t)$ and $\beta(t)$ obtained numerically using Eqs.~\eqref{eq:alpha_t_numerics} and~\eqref{eq:beta_t_numerics} (dashed lines), which agrees well with the exact result at any time. 
These figures show that the time evolution of the entanglement asymmetry qualitatively changes whether the Gaussian variational principle or the Bogoliubov theory is employed. 
In the former, the entanglement asymmetry increases in time, while in the latter, it decreases.
We also find that the analytic prediction~\eqref{eq:EA(t)} with Eqs.~\eqref{eq:alpha_t} and~\eqref{eq:beta_t} agrees well with the exact result for large-time scales, but it does not for short-time scales except for $t=0$. 
\par 
To better understand Fig.~\ref{fig:EA} (a), we plot separately the time evolution of $\alpha(t)$ and $\beta(t)$  in Figs.~\ref{fig:EA} (b) and (c), respectively. We can see in Fig.~\ref{fig:EA} (b) that the behavior of $\alpha(t)$ calculated with the Gaussian variational principle is qualitatively the same as that with the Bogoliubov theory (in the inset). 
Namely, it initially decreases with time and eventually tends to zero as predicted by Eq.~\eqref{eq:alpha_t}. 
On the other hand, the behavior of $\beta(t)$ drastically changes depending on which theory is employed to describe the initial state as shown in Fig.~\ref{fig:EA}~(c). 
In the Gaussian variational principle, $\beta(t)$ increases in time and finally converges to the stationary value given by Eq.~\eqref{eq:beta(inf)}. 
On the other hand, in the Bogoliubov theory (in the inset), $\beta(t)$ is almost constant in time and is much smaller than the one calculated with the Gaussian variational principle. 
We thus conclude that the qualitative difference in the entanglement asymmetry between the Gaussian variational principle and the Bogoliubov theory originates from the term~$\beta(t)$. 
\par 
In fact, according to Eqs.~\eqref{eq:beta_0} and~\eqref{eq:beta(inf)}, $\beta(t)$ contains the contribution of the condensate suppressed by an exponential term that only depends on the quantum depletions. This suppression is enhanced as the population of quantum depletions with long wavelength increases. 
Therefore, the smallness of $\beta(t)$ in the Bogoliubov theory may be due to the divergence of the mode occupation number of quantum depletions in the long-wavelength limit ${\bf k\to0}$ shown in Fig.~\ref{fig:n_k&m_k&n_0} (a). 
\par 

\section{Quantum fidelity}\label{sec:quantum fidelity}
As we already mentioned in the introduction, the entanglement asymmetry is usually employed to 
study how fast the subsystem reaches the stationary state using the local restoration of a symmetry broken by the initial state and respected by the post-quench Hamiltonian. We have just seen that the particle-number symmetry, spontaneously broken in the superfluid states, is not restored in a subsystem. Therefore, we cannot use the corresponding entanglement asymmetry to analyze the quantum Mpemba effect. As shown experimentally in Ref.~\cite{joshi-24}, this phenomenon should not be specific to a particular observable. For example, in addition to the entanglement asymmetry, that work also considers a distance between the time-evolved reduced density matrix $\rho_A(t)$ and its stationary value $\rho_A(\infty)$.
Therefore, in this section, we take the quantum fidelity between these states and investigate with it the occurrence of the quantum Mpemba effect in our quench. 

\subsection{Quantum fidelity and dimensional reduction}\label{subsec:fidelity}

The (logarithmic) quantum fidelity between the time-evolved reduced density matrix $\rho_A(t)$ and the stationary state $\rho_A(\infty)$ is defined as 
\begin{align}
    f_A(t)
    = 
    -
    \frac{1}{V_A}
    \log \Tr(\sqrt{\sqrt{\rho_A(t)}
    \rho_A(\infty)\sqrt{\rho_A(t)}}).
    \label{eq:f(t) definition}
\end{align}
This is a positive definite quantity, $f_A(t)\geq0$, that vanishes, $f_A(t)=0$, if and only if $\rho_A(t)=\rho_A(\infty)$. 
\par  
The quantum fidelity between two Gaussian states, such as $\rho_A(t)$ and $\rho_A(\infty)$ in our protocol, can be expressed in terms of their covariance matrices. Applying the results of Ref.~\cite{Banchi-2015}, we have in our case that
\begin{equation}\label{eq:f(t) exact}
f_A(t)=\frac{1}{4V_A}\log\det\left[\frac{\Gamma(t)+\Gamma(\infty)}{C^+(t)C^-(t)}\right],
\end{equation}
where 
\begin{align}
    C^+(t)
    &=
    2\Omega [\Gamma(t)+\Gamma(\infty)]^{-1}
    [\Omega/4+\Gamma(\infty)\Omega\Gamma(t)], 
    \\
    C^-(t)
    &=
    \sqrt{I+[C^+(t)\Omega]^{-2}/8}+I. 
\end{align}
Here $\Gamma(\infty)$ is the covariance matrix of $\rho_A(\infty)$. 
We can determine it from the one of $\rho_A(t)$, which we obtained in
Eq.~\eqref{eq:MV}. At large times $t$, the oscillating function $e^{-2\im t \xi_{\bf k}\sigma_y}$ in Eq.~\eqref{eq:g_k(t)} averages to zero, and $\Gamma(\infty)$ is given by Eq.~\eqref{eq:MV} with $g_\mathbf{k}(t)$ replaced by
\begin{align}
    g_{\bf k}(\infty)
    =
    \frac{\cosh(\theta_{\bf k})}{2}I.
    \label{eq:g_k(inf)}
\end{align}
\par 
As in the previous sections, we take as subsystem $A$ the strip of width $\ell$ represented in Fig.~\ref{fig:torus} and employ the dimensional reduction approach to calculate analytically the quantum fidelity~\eqref{eq:f(t) exact}.
Under the partial Fourier transformation~\eqref{eq:M}, the matrices appearing in Eq.~\eqref{eq:f(t) exact} take the form
\begin{equation}\label{eq:Gamma block diagonal}
M\Gamma(t)M=\bigoplus_{k_y}\Gamma_{k_y}(t),
\end{equation}
where
\begin{equation}
\Gamma_{k_y}(t)=\frac{1}{L_x}\sum_{k_x}e^{\im k_x(i_x-i_x')}g_{\mathbf{k}}(t),
\end{equation}
and
\begin{align}
    MC^\pm(t)M^\dag 
    &=
    \bigoplus_{k_y}C_{k_y}^\pm(t),
    \label{eq:C block diagonal}
 \end{align}
with 
\begin{align}
    C_{k_y}^+(t)
    &=
    2\tilde \Omega[\Gamma_{k_y}(t)+\Gamma_{k_y}(\infty)]^{-1}
    \nonumber\\
    &\quad \quad \quad 
    [\tilde \Omega/4+\Gamma_{k_y}(\infty)\tilde \Omega\Gamma_{k_y}(t)],
    \label{eq:C_k}
    \\
     C_{k_y}^-(t)
    &=
    \sqrt{I+[C_{k_y}^+(t)\tilde \Omega]^{-2}/8}+I, 
    \label{eq:tilde C_k}
\end{align}
and
\begin{align}
    \tilde \Omega=\bigoplus_{i=1}^{\ell} \mqty(0&1\\-1&0).
\end{align}
The block structure of $\Gamma(t)$ and $C^\pm(t)$ in Eqs.~\eqref{eq:Gamma block diagonal} and \eqref{eq:C block diagonal} allows us to decompose the quantum fidelity \eqref{eq:f(t) exact} into the contribution of each transverse momentum sector as 
\begin{align}
    f_A(t)
    =
    \sum_{k_y}f_{k_y}(t),
    \label{eq:f(t)=sum_k f_k(t)}
\end{align}
with 
\begin{align}
    f_{k_y}(t)
    &=
    \frac{1}{4V_A}
    \log\det\left[\frac{\Gamma_{k_y}(t)+\Gamma_{k_y}(\infty)}{C_{k_y}^+(t)C_{k_y}^-(t)}\right].
    \label{eq:f_k(t)}
\end{align}

Note that the decomposition above is exact for any subsystem size. 
We will numerically calculate the quantum fidelity with Eqs.~\eqref{eq:f(t)=sum_k f_k(t)} and~\eqref{eq:f_k(t)} and use the result as a benchmark for the analytic predictions reported in the following section. 

\subsection{Time evolution of the quantum fidelity and the quantum Mpemba effect}\label{subsec:time_evolution_fidelity}

We can now examine the time evolution of the quantum fidelity~\eqref{eq:f(t) definition} after the quench to the free Hamiltonian from the ground state of the Bose-Hubbard model deep in the superfluid phase, using both the Bogoliubov theory and the Gaussian variational approach for the initial state.

In Appendix~\ref{apdx:mult_st_app}, we calculate analytically the quantum fidelity in the ballistic regime $\ell, t\to \infty$, with $\zeta=t/\ell$ fixed, using the decomposition~\eqref{eq:f(t)=sum_k f_k(t)} in the transverse momentum
sectors and the multidimensional stationary phase approximation. We obtain that  
\begin{align}
    f_A(t)
    \simeq 
    \frac{1}{L_y}
    \sum_{k_y}
    \int_{-\pi}^\pi
    \frac{{\rm d}k_x}{8\pi}
    x_{k_x}(\zeta)\log(1+3\abs{\tilde m_{\bf k}}^2).
    \label{eq:f(t) analytic}
\end{align}
Once again, this result can be interpreted in terms of the quasiparticle picture for quantum depletions. The function $x_{k_x}(\zeta)$ counts the number of pairs of quantum depletions with opposite momenta inside the subsystem and $\tilde m_{\bf k}=\bra{\rm SF}b_{\bf k}b_{-\bf k}\ket{\rm SF}$ is the anomalous correlation function between the quantum depletions forming a pair. As the number of pairs with both excitations inside $A$ reduces in time, $f_A(t)$ decreases and vanishes in the limit $t\to\infty$. That is, $\rho_A(t)=\rho_A(\infty)$ when the number of complete pairs of quantum depletions in the subsystem is zero.
\par 
\begin{figure}
    \raggedright
    \includegraphics[width=0.9\linewidth]{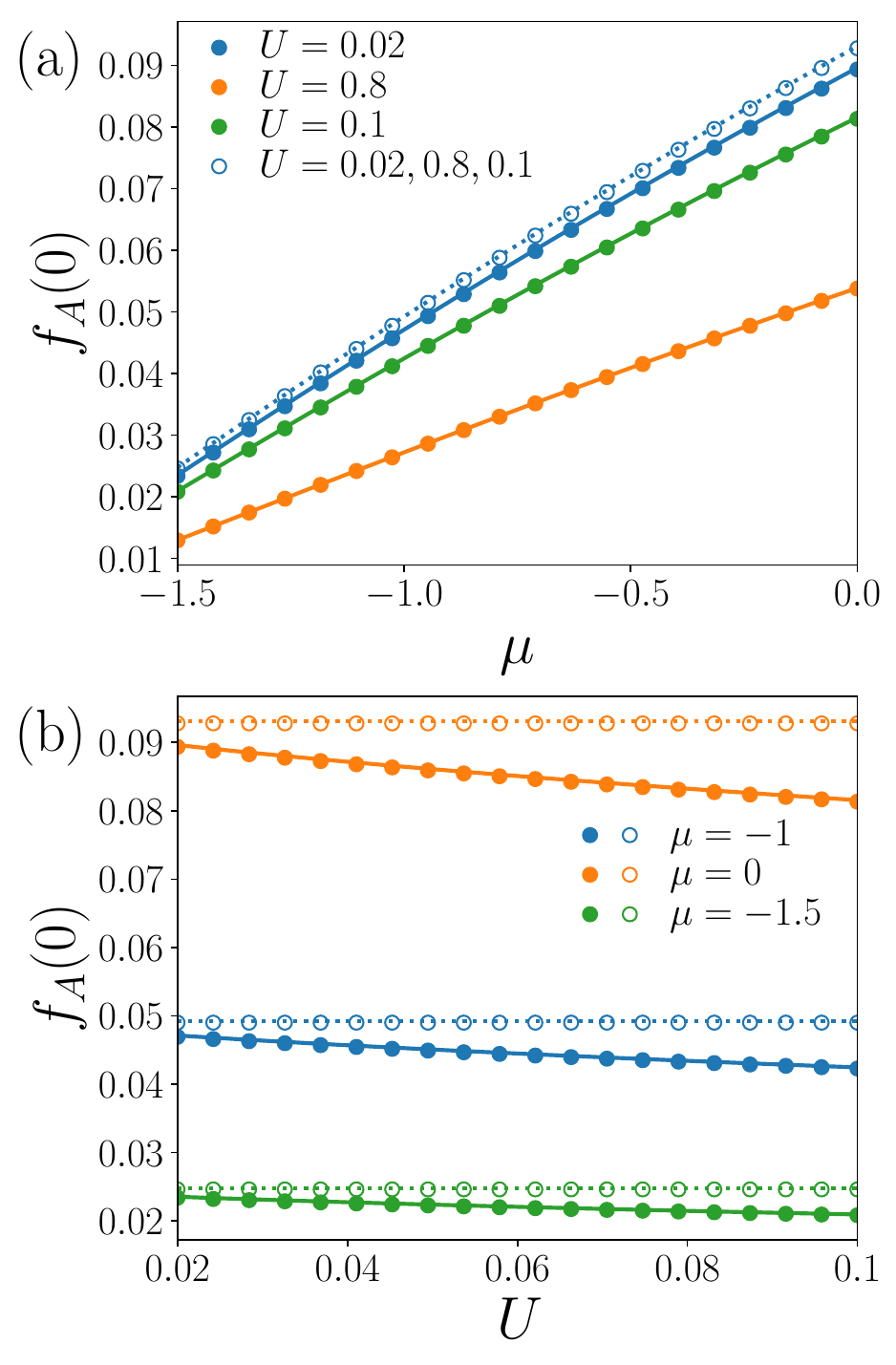}
    \caption{Quantum fidelity in the superfluid state as a function of the chemical potential $\mu$ for a fixed interaction strength $U$ (upper panel) and vice-versa (lower panel). The curves and the symbols correspond to the analytic prediction~\eqref{eq:f(t) analytic} and to the exact numerical values obtained using Eq.~\eqref{eq:f(t) exact}, respectively. 
    The filled and empty symbols correspond to taking the Gaussian variational principle and the Bogoliubov theory, respectively, to describe the initial state. 
    We take $\ell=200$ and $L_y=50$ in all the plots.}
    \label{fig:fidelity_t0}
\end{figure}
In Figs.~\ref{fig:fidelity_t0} (a) and (b), we plot $f_A(t)$ at \( t = 0 \) as a function of \( \mu \) for fixed \( U \) and vice-versa, respectively. The plots show that the initial superfluid state is farther from the corresponding stationary state increasing \( \mu \) and decreasing \( U \) in the Gaussian variational principle. The same behavior is observed in the Bogoliubov theory when varying $\mu$, while $f_A(0)$ is constant in $U$ in this case. This trend reflects the fact that the anomalous correlations between quantum depletions, \( \tilde{m}_{\bf k} \), which are the only contribution to the quantum fidelity according to Eq.~\eqref{eq:f(t) analytic}, grow as \( \mu \) increases and \( U \) decreases, as we can also see in the panel (b) of Fig.~\ref{fig:fidelity_t0}.
\par
\begin{figure}
    \raggedright
    \includegraphics[width=0.49\textwidth]{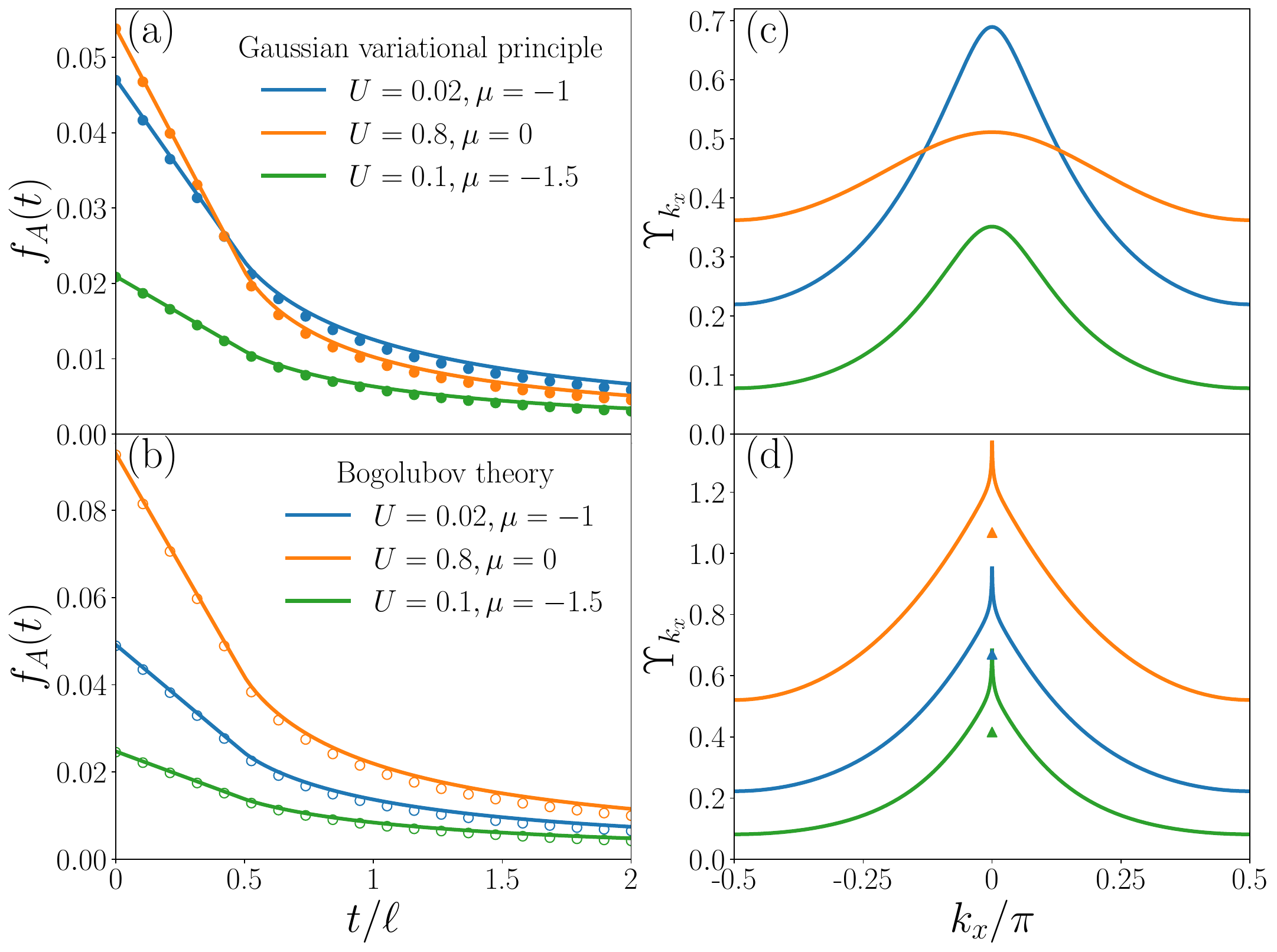}
    \caption{Left panels: Time evolution of the quantum fidelity after the quench to the free system from different superfluid states. The solid lines and the symbols correspond to the analytic prediction \eqref{eq:f(t) analytic} and the exact numerical result obtained by using Eq.\,\eqref{eq:f(t) exact}, respectively. In panel (a), the initial
    superfluid state is obtained using the 
    Gaussian variational principle while in panel (b) we employ the Bogoliubov theory.
    Right panels: Contribution $\Upsilon_{k_x}$ of the quantum depletion pairs to the fidelity as a function of the longitudinal momenta $k_x$ in the Gaussian variational approach (panel (c)) and in the Bogoliubov theory (panel (d)). 
    The symbols in panel (d) represent the value of $\Upsilon_{k_x}$ at $k_x=0$ for each set of parameters $U$ and $\mu$. We take $\ell=200$ and $L_y=50$ in all the plots.
    }
    \label{fig:QuantumFidelity}
\end{figure}

 
In Figs.~\ref{fig:QuantumFidelity}~(a)~and~(b), we plot the quantum fidelity after the quench as a function of time for several initial superfluid states with different chemical potential $\mu$ and repulsive interaction $U$. In panel (a), we use the Gaussian variational principle to obtain the superfluid state while in panel (b) we apply the Bogoliubov theory for the same values of $\mu$ and $U$. In both cases, the analytic prediction~\eqref{eq:f(t) analytic} (solid curves) agrees well with the exact result (symbols), computed numerically using Eq.~\eqref{eq:f(t) exact}.
As expected, $f_A(t)$ decreases in time and finally tends to zero at large times, indicating the relaxation of the subsystem to the stationary state $\rho_A(\infty)$. 
We find that, for certain pairs of initial states, their quantum fidelities cross at a particular time when applying the Gaussian variational principle. This means that the system initially farther from the stationary state relaxes to it faster, signaling the occurrence of the quantum Mpemba effect.
On the contrary, when we apply the Bogoliubov theory, the quantum fidelities of any pair of initial states never cross during the time evolution, implying that the quantum Mpemba effect is absent in this approach.  

\subsection{Microscopic conditions for the quantum Mpemba effect}\label{subsec:condition}

To elucidate the reasons behind the phenomenology of Figs.~\ref{fig:QuantumFidelity} (a) and (b), here we investigate the conditions for the quantum Mpemba effect to occur in the present setup. 
\par 
Let us consider the quench to the free Hamiltonian~\eqref{eq:H_free} from two different initial superfluid states, $\ket{\rm SF_1}$ and $\ket{\rm SF_2}$.
If subsystem $A$  is farther from its stationary state in $\ket{\rm SF_1}$, then their quantum fidelities must satisfy  
\begin{align}
    f_{A,1}(0)>f_{A,2}(0), 
    \label{ineq:f_1(0)>f_2(0)}
\end{align}
where $f_{A,i}(t)$ is the quantum fidelity for the state $i$ ($i=1,2$). 
In this situation, the quantum Mpemba effect occurs when there exists a time $t_{\rm M}$ after which the subsystem $A$ in the state 1 is closer to the corresponding stationary state than in the state 2, i.e., 
\begin{align}
    f_{A,1}(t)<f_{A,2}(t)\ \forall t>t_{\rm M}. 
    \label{ineq:f_1(t)<f_2(t)}
\end{align}
\par 
We can write Inequalities~\eqref{ineq:f_1(0)>f_2(0)} and~\eqref{ineq:f_1(t)<f_2(t)} in terms of
the microscopic properties of the initial states, i.e. their density of occupied modes or their anomalous correlations, using the analytic expression obtained in Eq.~\eqref{eq:f(t) analytic} for $f_A(t)$. To this end, it is convenient to rewrite it as follows 
\begin{align}
    f_A(t)
    = 
    \int_{-k^*_x(t)}^{k^*_x(t)}
    \frac{{\rm d}k_x}{2\pi}
    (1-2\zeta|v_{k_x}|) \Upsilon_{k_x}
    \label{eq:f(t) rewrite}
\end{align}
where 
\begin{align}
    k_x^*(t)
    =
    \arcsin(\frac{\ell}{2t})\Theta(2t-\ell)
    +\frac{\pi}{2}\Theta(\ell-2t),
\end{align}
with $\Theta(x)$ being the Heaviside step function, and
\begin{multline}
    \Upsilon_{k_x}
    = 
    \frac{1}{4L_y}
    \sum_{k_y}
    \Big[\log(1+3\abs{\tilde m_{(k_x,k_y)}}^2)
    \\
    +\log(1+3\abs{\tilde m_{(k_x+\pi,k_y)}}^2)\Big].
\end{multline}
Observe that we have replaced the function 
$x_{k_x}(\zeta)$ in Eq.~\eqref{eq:f(t) analytic}, which counts the number of entangled pairs of quantum depletions inside the subsystem, by a modified domain of integration $k_x\in [-k_x^*(t),k_x^*(t)]$ in the expression~\eqref{eq:f(t) rewrite}.
As $t$ increases, this integration domain becomes narrower around $k_x=0$ and, finally, it vanishes in the large time limit $t\to\infty$. 
\par 
Applying Eq.~\eqref{eq:f(t) rewrite}, the $t=0$ condition~\eqref{ineq:f_1(0)>f_2(0)} can be written as 
\begin{align}
    \int_{-\pi/2}^{\pi/2}
    {\rm d}k_x
    \Upsilon_{k_x,1}
    >
    \int_{-\pi/2}^{\pi/2}
    {\rm d}k_x
    \Upsilon_{k_x,2}, 
    \label{ineq:1}
\end{align}
where $\Upsilon_{k_x,i}$ is $\Upsilon_{k_x}$ for the state $i$ ($i=1,2$). 
This result implies that, at $t=0$, all the entangled pairs in the quantum depletion contribute to the fidelity, being their total contribution larger in the configuration initially farther from the corresponding post-quench stationary state. 
\par 
On the other hand, Eq.~\eqref{eq:f(t) rewrite} allows us to rewrite inequality~\eqref{ineq:f_1(t)<f_2(t)} as follows 
\begin{align}
    &\int_{-k_x^*(t)}^{k_x^*(t)}
    \frac{{\rm d}k_x}{2\pi}
    (1-2\zeta|v_{k_x}|)
    \Upsilon_{k_x,1}
    \nonumber\\
    &\quad <
    \int_{-k_x^*(t)}^{k_x^*(t)}
    \frac{{\rm d}k_x}{2\pi}
    (1-2\zeta|v_{k_x}|)
    \Upsilon_{k_x,2}\ \forall t>t_{\rm M}.
    \label{ineq:f_1(t)<f_2(t)_2}
\end{align}
Since the integrands in the above inequality are positive definite and even for $k_x$, there always exists a large enough time $t_{\rm M}$ such that the inequality~\eqref{ineq:f_1(t)<f_2(t)_2} can be re-expressed as 
\begin{align}
    \Upsilon_{k_x,1}
    < 
    \Upsilon_{k_x,2}
    \ \forall k_x\in[-k_x^*(t_{\rm M}),k_x^*(t_{\rm M})]. 
    \label{ineq:2}
\end{align}
Given that $k_x^*(t_{\rm M})$ is small for large $t_{\rm M}$ as $k_x^*(t_{\rm M})\simeq \ell/(2t_{\rm M})$, the above inequality implies that the contribution of quantum depletions with the slowest longitudinal group velocity ($v_{k_x}\simeq0$) is smaller in the state 1 than in the state 2. 
\par 
Inequalities \eqref{ineq:1} and \eqref{ineq:2} are the sufficient and necessary conditions for the quantum Mpemba effect to be observed in our setup. 
According to them, this phenomenon occurs if, in the initial state that is farther from the post-quench stationary state in the subsystem $A$, the contribution to the fidelity of the quantum depletion modes with the slowest longitudinal group velocity $v_{k_x}$ is smaller than in the other configuration, initially closer to the equilibrium. 
\par 
In the right panels of Fig.~\ref{fig:QuantumFidelity}, we plot $\Upsilon_{k_x}$ as a function of $k_x$ for each set of values of $U$ and $\mu$ considered in the corresponding panel on the left. 
These figures show that $\Upsilon_{k_x}$ satisfies the microscopic conditions~\eqref{ineq:1} and~\eqref{ineq:2} for the pair of initial states whose quantum fidelities intersect at a finite time, i.e., the quantum Mpemba effect occurs. In contrast, these inequalities are not satisfied for the pairs of initial states for which the quantum Mpemba effect does not happen. 
\par 
In closing this section, we show that the conditions~\eqref{ineq:1} and~\eqref{ineq:2} are never met, i.e., the quantum Mpemba effect never occurs, in the Bogoliubov theory. To this end, let us consider two superfluid initial states with chemical potential $\mu_1$ and $\mu_2$, respectively. Since the anomalous correlation $\tilde m_{\bf k}$ does not depend on $U$ in the Bogoliubov theory and $\tilde m_{\bf k}$ is the only information about the initial state that enters in the conditions, the value of $U$ is irrelevant for the present discussion. 
If we denote as $\tilde m_{{\bf k},i}$ the anomalous correlation of the initial state $i$ ($i=1,2$), then the combination of Eqs.~\eqref{eq:Bogoliubov angle} and~\eqref{eq:m_k tilde} give 
\begin{align}
    \abs{\tilde m_{{\bf k},i}}^2
    = 
    \frac{\bar \mu^2_i}{16\bar \epsilon_{\bf k}(\bar \mu_i + \bar \epsilon_{\bf k})},
\end{align}
where 
\begin{gather}
    \bar \mu_i = 2+\mu_i>0,\\ 
    \bar \epsilon_{\bf k}
    = 
    2-\cos k_x-\cos k_y>0.
\end{gather}
Assuming $\mu_1>\mu_2$, for ${\bf k\neq0}$, we have
\begin{align}
    &
    \abs{\tilde m_{{\bf k},1}}^2
    -
    \abs{\tilde m_{{\bf k},2}}^2>0
\end{align}
While for ${\bf k=0}$, since we set $\theta_{\bf k}=0$, we have 
\begin{align}
    \abs{\tilde m_{{\bf 0},1}}^2
    =
    \abs{\tilde m_{{\bf 0},2}}^2
    =0.
\end{align}
We thus obtain 
\begin{align}
    \mu_1>\mu_2 
    &\Rightarrow 
    \abs{\tilde m_{{\bf k},1}}^2
    \geq 
    \abs{\tilde m_{{\bf k},2}}^2. 
\end{align}
This implies that 
\begin{align}
    \mu_1>\mu_2
    \Rightarrow 
    \Upsilon_{k_x,1}
    > \Upsilon_{k_x,2}
    \ \forall k_x. 
\end{align}
Therefore, in the Bogoliubov theory, the curves corresponding to $\Upsilon_{k_x}$ for different initial configurations never intersect. In fact, see in the panel (d) of Fig.~\ref{fig:QuantumFidelity} that the effect of modifying $\mu$ is simply a global vertical shift in $\Upsilon_{k_x}$.
This means that the conditions~\eqref{ineq:1} and ~\eqref{ineq:2} are never satisfied simultaneously and, therefore, the quantum Mpemba effect never occurs in this case. The effect of the on-site repulsion $U$ must be included in the anomalous correlations to observe it.

\section{Conclusions}\label{sec:conclusion}
In this paper, we investigated different aspects of the non-equilibrium dynamics of bosons in a two-dimensional periodic lattice. We focused on the quench from the superfluid phase of the Bose-Hubbard model to the free-boson system. To describe the initial configuration, we applied the Bogoliubov theory and a recently proposed extension, the Gaussian variational principle. 
These two methods offer the advantage of approximating the ground state of the superfluid phase as a Gaussian state.
Since the free time evolution preserves the Gaussianity of the state, we can perform exact computations. 
Using both approximations, we analyzed the time evolution of the entanglement entropy, the entanglement asymmetry of the particle-number symmetry (spontaneously broken by the initial state), and the occurrence of the quantum Mpemba effect.
Our analysis focused on a subsystem in the form of a strip extended along one direction.
This choice allowed us to apply dimensional reduction and effectively transform the 2D system into decoupled 1D chains. 
Exploiting then well-known techniques for 1D free systems, we obtained exact analytic results in the ballistic regime. Our analysis reveals that, in general, the Bogoliubov theory and the Gaussian variational principle lead to qualitatively different behaviors of the system following the quench.
\par 
We examined the entanglement following the quench by calculating the R\'enyi entanglement entropies. Our results show that their time evolution can be explained using the standard quasiparticle picture, where the propagation of quantum depletions in entangled pairs drives the entanglement growth after the quench.
In particular, the entanglement entropy in the stationary state is determined by the mode occupancy of quantum depletions in the initial superfluid state. This implies that, in the Bogoliubov theory, the R\'enyi entanglement entropies are independent of the on-site repulsive interaction $U$ because interactions between quantum depletions are neglected. 
In contrast, within the Gaussian variational principle, the entropies depend on $U$ since the interactions between quantum depletions affect their mode occupancy.
\par 
To study the dynamics of the spontaneously broken U(1) particle-number symmetry after the quench, we analyzed the corresponding entanglement asymmetry.  Traditionally, the breaking of this symmetry is detected with the superfluid order parameter. 
However, we have found that the entanglement asymmetry provides more comprehensive information about the symmetry breaking, as it incorporates the effect of both the Bose-Einstein condensate and the quantum depletions. 
When the initial state is approximated using the Gaussian variational principle, we observe an increase in the entanglement asymmetry after the quench. 
In contrast, employing the Bogoliubov theory results in a decrease.
In both cases, the contribution from quantum depletions, which is captured by the quasiparticle picture, vanishes at large times. However, there is a non-zero contribution from the condensate. 
Therefore, even though the post-quench dynamics respects the symmetry, the entanglement asymmetry tends to a non-zero value, indicating the lack of local symmetry restoration due to the presence of the condensate. 
The qualitative difference between the two approximations of the initial state in the dynamics of the entanglement asymmetry stems from the divergence of the mode occupancy of quantum depletions in the Bogoliubov theory at long wavelengths, which strongly affects the contribution of the condensate to the asymmetry.  
\par 
Lastly, we investigated the subsystem relaxation and the occurrence of the quantum Mpemba effect. In the prior literature, this was typically assessed by monitoring the local restoration of an initially broken symmetry using entanglement asymmetry. However, since the particle-number symmetry is not restored in our case, we opted to measure the distance between the time-evolved reduced density matrix and the corresponding stationary state using their quantum fidelity.
We derived its exact time evolution in the ballistic regime and showed that is described by the quasiparticle picture for the quantum depletions. 
Using the Gaussian variational principle, we have seen that there are pairs of superfluid states for which the configuration initially farther from its stationary state relaxes to it faster, signaling the occurrence of the quantum Mpemba effect. We derived the conditions for the appearance of this effect in terms of the mode occupancy of quantum depletions in the initial state. We further showed that they are never satisfied if we use the Bogoliubov theory to describe the initial state. 
\par 
One of our main findings is the distinct qualitative behaviors of observables when approximating the initial state using Bogoliubov theory versus the Gaussian variational principle. Since the latter is expected to provide a more accurate description of the initial state, we can conclude that in the actual Bose-Hubbard model, we should observe the quantum Mpemba effect and an increase of entanglement asymmetry, in contrast with the predictions of the Bogoliubov approximation.    
\par
The effective theories employed here to characterize the initial superfluid state assume that bosons are weakly interacting. However, exploring the strongly interacting regime would also be compelling. For example, combining our methods with the slave boson approach~\cite{dvods-03, fr-16} or the Gutzwiller approximation~\cite{pekker-12} could facilitate the study of the quench dynamics between the Mott-insulating and superfluid phases. Extending our analysis to spin-1 bosons in optical lattices~\cite{dz-02, tkk-04, kt-13, mt-13} is another intriguing direction, as the superfluid phase in this system displays a variety of magnetic patterns and symmetries. Such quench dynamics could uncover a rich physics arising from the interplay between the internal degrees of freedom of bosons and the strong correlations among them.
\acknowledgements
We thank C. Rylands for the fruitful discussions. 
SY is supported by Grant-in-Aid for JSPS Fellows (Grant No. JP22J22306). PC and FA acknowledge support from ERC under Consolidator Grant number 771536 (NEMO) and from European Union - NextGenerationEU, in the framework of the PRIN Project HIGHEST number 2022SJCKAH\_002. 

\appendix

\section{Multidimensional stationary phase approximation}\label{apdx:stationary_phase}

In this Appendix, we describe in detail the multidimensional 
stationary phase approach following Refs.~\cite{fc-08, cef-12}
that we apply to derive many of the analytic results of the main text. 

In general, we need to evaluate the moments ${\rm Tr}(T^m)$ of a $2\ell\times 2\ell$ block-Toeplitz matrix, say $T$, 
\begin{equation}
T_{jj'}=\int_{0}^{2\pi}\frac{{\rm d}k}{2\pi}e^{-\im (j-j')k}g(k, t),
\end{equation}
generated by a $2\times 2$ symbol of the form   
\begin{align}
    g(k, t)=c_kI
    +
    d_k\sigma_z e^{-2\im t \xi_k \sigma_y}, 
\end{align}
where $c_k$ and $d_k$ are time-independent functions. 
The moments of $T$ can be written as 
\begin{multline}
    \Tr_{}(T^m)
    = 
    \sum_{i_1=1}^\ell 
    \cdots 
    \sum_{i_m=1}^\ell 
    \int_{[-\pi,\pi]^m}
    \frac{{\rm d}^m\boldsymbol{k}}{(2\pi)^m}
    \\
    \times e^{-\im \sum_{j=1}^m k_j(i_{j}-i_{j+1})}
    F(\boldsymbol{k}), 
    \label{eq:moments_of_T}
\end{multline}
where 
\begin{align}
    F(\boldsymbol{k})
    =
    \Tr(\prod_{i=1}^m(c_{k_i}I+d_{k_i}\sigma_z e^{-2\im t \xi_{k_i} \sigma_y})).
\end{align}
Using the identity, 
\begin{align}
    \sum_{i=1}^\ell 
    e^{-\im ki}
    =
    e^{-\frac{\im}{2}(\ell+1)k}
    \frac{\ell}{2}
    \int_{-1}^1 
    {\rm d}u
    \frac{k}{2\sin(\frac{k}{2})}
    e^{-\frac{\im }{2}\ell k u}, 
    \label{eq:sum_to_integral}
\end{align}
the sums over $i_{j}$, $j=1,\dots, m$, in Eq.~\eqref{eq:moments_of_T} can be replaced by the integrals over $u_{j}$,  
\begin{multline}
    \Tr_{}(T^m)
    =
    \qty(\frac{\ell}{2})^{m}
    \int_{[-1,1]^m}
    {\rm d}^m\boldsymbol{u}
    \int_{[-\pi,\pi]^m}
    \frac{{\rm d}^m\boldsymbol{k}}{(2\pi)^m}
    \\
    \times e^{-\frac{\im}{2}\ell\sum_{i=1}^m k_i(u_i-u_{i+1})}
    C(\boldsymbol{k}) F(\boldsymbol{k}), 
    \label{eq:moment_integral}
\end{multline}
with 
\begin{align}\label{eq:C_k}
    C(\boldsymbol{k})
    =
    \prod_{i=1}^{m}
    \frac{k_i-k_{i+1}}{2\sin(\frac{k_i-k_{i+1}}{2})}.
\end{align}
Introducing the new variables 
\begin{align}
    z_i 
    =
    \begin{dcases}
        u_1&i=1,\\
        u_{i+1}-u_i&i=2,...,m,
    \end{dcases}
\end{align}
Eq.~\eqref{eq:moment_integral} can be rewritten as 
\begin{multline}
    \Tr_{}(T^m)
    =
    \qty(\frac{\ell}{2})^{m}
    \int_{R_z}
    {\rm d}^m\boldsymbol{z}
    \int_{[-\pi,\pi]^m}
    \frac{{\rm d}^m\boldsymbol{k}}{(2\pi)^m}
    \\
    \times e^{-\frac{\im}{2}\ell\sum_{i=2}^m z_i(k_i-k_1)}
    C(\boldsymbol{k}) F(\boldsymbol{k}), 
    \label{eq:moment_integral_2}
\end{multline}
where the integral domain $R_z$ is 
\begin{align}
    R_z:~\sum_{i=1}^{j}z_i\in[-1,1]\quad \forall j\in\{1,2,...,m\}.
\end{align}
Since the integrand in Eq.~\eqref{eq:moment_integral_2} is independent of $z_1$, the integral over it can be easily calculated. We find
\begin{multline}
    \Tr_{}(T^m)
    =
    \qty(\frac{\ell}{2})^{m}
    \int_{\mathbb{R}^{m-1}}
    {\rm d}^{m-1}\boldsymbol{z}
    \int_{[-\pi,\pi]^m}
    \frac{{\rm d}^m \boldsymbol{k}}{(2\pi)^m}
    \\
    \times e^{-\frac{\im}{2}\ell\sum_{i=2}^m z_i(k_i-k_1)}
    \nu(\boldsymbol{z})
    C(\boldsymbol{k}) F(\boldsymbol{k}), 
    \label{eq:moment_integral_3}
\end{multline}
with 
\begin{align}
    \nu(\boldsymbol{z})
    =
    \max\qty(
    0,
    2-\max_{1\leq j\leq m}
    \qty(\sum_{i=2}^jz_i)
    +
    \min_{1\leq j\leq m}
    \qty(\sum_{i=2}^jz_i)
    ). 
\end{align}
Expanding $F(\boldsymbol{k})$, the right-hand side of Eq.~\eqref{eq:moment_integral_3} can be written as 
\begin{widetext}
\begin{multline}
    \Tr(T^m)
    =
    2
    \qty(\frac{\ell}{2})^{m}
    \int_{\mathbb{R}^{m-1}}
    {\rm d}^{m-1}\boldsymbol{z}
    \int_{[-\pi,\pi]^m}
    \frac{{\rm d}^m\boldsymbol{k}}{(2\pi)^m}e^{-\frac{\im}{2}\ell\sum_{i=2}^m z_i(k_i-k_1)}
    \nu(\boldsymbol{z})
    C(\boldsymbol{k})
    \\
    \times \sum_{p=0}^{\lfloor \frac{m}{2} \rfloor}
    \sum_{1\leq j_1<j_2,...,j_{2p}\leq m}
    \qty(\prod_{i\notin\{j_1,..,j_{2p}\}} c_{k_i})
    \qty(\prod_{i=1}^{2p}d_{k_{j_i}})
    \cos(2t\sum_{i=1}^{2p}(-1)^{i}\xi_{k_{j_i}}).
    \label{eq:moment_integral_4}
\end{multline}
\end{widetext}
Since the integral is invariant under the permutations $z_{j_i}\leftrightarrow z_i$ and $k_{j_i}\leftrightarrow k_i$, Eq.~\eqref{eq:moment_integral_4} can be rewritten as 
\begin{align}
    \Tr(T^m)
    =
    \qty(\frac{\ell}{2})^{m}
    \int_{-\pi}^\pi 
    \frac{{\rm d}k_1}{2\pi}
    \sum_{p=0}^{\lfloor \frac{m}{2}\rfloor}
    \mqty(m\\2p)
    \mathcal{I}_{2p}(k_1), 
    \label{eq:moment_integrand_5}
\end{align}
where 
\begin{multline}
    \mathcal{I}_{2p}(k_1)
    =
    \int_{\mathbb{R}^{m-1}}
    {\rm d}^{m-1} \boldsymbol{z}
    \int_{[-\pi,\pi]^{m-1}} 
    \frac{{\rm d}^{m-1}\boldsymbol{k}}{(2\pi)^{m-1}}
    C(\boldsymbol{k})\nu(\boldsymbol{z})
    \\
    \times\qty(\prod_{i=2p+1}^{m}c_{k_i})
    \qty(\prod_{i=1}^{2p} d_{k_i})
    e^{\im \ell \vartheta(\boldsymbol{k},\boldsymbol{z})}
    +{\rm c.c.}, 
    \label{eq:I_2p(k_1)}
\end{multline}
with 
\begin{align}
    \vartheta(\boldsymbol{k},\boldsymbol{z})
    =
    \frac{1}{2}
    \sum_{i=2}^m 
    z_i(k_i-k_1)
    -
    2\zeta\sum_{i=1}^{2p}(-1)^{i}\xi_{k_{i}}.
\end{align}
\par 
Observing that the phase factor of the integrand in $\mathcal{I}_{2p}(k_1)$ rapidly oscillates for $\ell\gg1$, we can evaluate its asymptotic form in the ballistic limit by applying the multidimensional stationary phase approximation. It states that, for $\ell\gg 1$, the $N$-dimensional oscillating integral behaves as
\begin{multline}
    \int {\rm d}^N {\bf x}
    f(\mathbf{x})
    e^{\im \ell g(\mathbf{x})}
    \simeq
    \qty(\frac{2\pi}{\ell})^{N/2}
    \sum_{\bar{\bf x}}
    f(\bar {\mathbf{x}})
    \abs{{\rm det}{\rm H}(g(\bar{\bf x}))}^{-\frac{1}{2}}\\
    \times 
    e^{\im \ell g(\bar{\mathbf{x}})
    +\frac{\im}{4}\pi {\rm sgn}({\rm H}(g(\bar{\bf x})))}.
    \label{eq:formula_stationary_phsae}
\end{multline}
Here $\bar{\bf x}$ is the stationary point of $g(\mathbf{x})$ at which $\nabla g(\bar{\bf x})=0$, ${\rm H}(g({\bf x}))$ is the Hessian matrix of $g({\bf x})$, whose entries are given by ${\rm H}(g({\bf x}))_{ii'}=\partial_{x_i}\partial_{x_{i'}}g(\mathbf{x})$, and ${\rm sgn}({\rm H}(g(\bar{\bf x})))$ stands for the sign of the difference between the positive and negative eigenvalues of ${\rm H}(g(\bar{\bf x}))$.
The stationary points of $\mathcal{I}_{2p}(k_1)$ are obtained by solving the equation $\nabla \vartheta(\bar{\boldsymbol{k}},\bar{\boldsymbol{z}})=0$ and read 
\begin{align}
    \bar k_i =k_1,~
    \bar z_i
    = 
    \begin{dcases}
        4\zeta(-1)^i v_{k_1} &i=2, \dots, 2p,\\
        0 & i>2p.
    \end{dcases}
\end{align}
At the stationary points, we have 
\begin{gather}
    C(\bar{\boldsymbol{k}})\qty(\prod_{i=2p+1}^{m}c_{\bar k_i})
    \qty(\prod_{i=1}^{2p} d_{\bar k_i})
    =c_{k_1}^{m-2p}
    d_{k_1}^{2p},
     \label{eq:stationary point1}
    \\
    \nu(\bar{\boldsymbol{z}})
    = 
    \begin{dcases}
        2&p=0\\
        2x_{k_1}(\zeta)&p>0
    \end{dcases},
      ~\vartheta(\bar{\boldsymbol{k}},\bar{\boldsymbol{z}})=0,
    \label{eq:stationary point2}
\end{gather}
and
\begin{gather}
    |{\rm det H}(\vartheta(\bar{\boldsymbol{k}},\bar{\boldsymbol{z}}))|
    =
    4^{1-m},~
    {\rm sgn}({\rm H}(\vartheta(\bar{\boldsymbol{k}},\bar{\boldsymbol{z}})))=0.
    \label{eq:stationary point3}
\end{gather}
Applying the stationary phase approximation~\eqref{eq:formula_stationary_phsae} to Eq.~\eqref{eq:I_2p(k_1)} and using Eqs.~\eqref{eq:stationary point1},~\eqref{eq:stationary point2},~and~\eqref{eq:stationary point3}, we obtain 
\begin{align}
    \mathcal{I}_{2p}(k_x)
    \simeq 
    \qty(\frac{2}{\ell})^{m-1} 
    c_{k_x}^{m-2p}
    d_{k_x}^{2p}
    \times 
    \begin{cases}
        2&p=0\\
        2x_{k_x}(\zeta) &p>0
    \end{cases}.
    \label{eq:I_2p stationary}
\end{align}
Here we rewrote $k_1$ as $k_x$. Inserting Eq.~\eqref{eq:I_2p stationary} into Eq.~\eqref{eq:moment_integrand_5}, we arrive at 
\begin{align}
    \Tr_{}(T^m)
    &\simeq 
    2\ell 
    \int_{-\pi}^\pi 
    \frac{{\rm d}k_x}{2\pi}
    c_{k_x}^m 
    \nonumber\\
    &+
    2\ell 
    \int_{-\pi}^\pi 
    \frac{{\rm d}k_x}{2\pi}
    x_{k_x}(\zeta) \sum_{p=1}^{\lfloor \frac{m}{2}\rfloor}
    \mqty(m\\2p)
    c_{k_x}^{m-2p}d_{k_x}^{2p},
    \nonumber
\end{align}
and, performing the sum in the integrand,
\begin{align}
     \Tr_{}(T^m)&\simeq 
    2\ell 
    \int_{-\pi}^\pi 
    \frac{{\rm d}k_x}{2\pi}
    [1-x_{k_x}(\zeta)]c_{k_x}^{m}
    \nonumber\\
    &+
    \ell 
    \int_{-\pi}^\pi
    \frac{{\rm d}k_x}{2\pi}
    x_{k_x}(\zeta)
    [(c_{k_x}+d_{k_x})^m+(c_{k_x}-d_{k_x})^m].
    \label{eq:Tr(T^m)_final}
\end{align}

\section{Derivation of Eq.~\eqref{eq:charged moment exact}}\label{apdx:charged_moment}
Here we derive Eq.~\eqref{eq:charged moment exact} using the Wigner-function formalism for bosonic Gaussian states, see e.g.~\cite{bvl-05}. 
\par 
Let us first introduce the coherent state $\ket{{\bf x}}$, this is the eigenstate of the canonical operator $x_{\bf i}$ with eigenvalue ${\rm x}_{\bf i}$ $({\bf i}\in A)$. 
The Wigner function of the reduced density matrix $\rho_A$ is given by 
\begin{align}
    W({\bf x,p})
    =
    \int_{\mathbb{R}^{V_A}}
    \frac{{\rm d}{\bf q}}{(2\pi)^{V_A}}
    \bra{{\bf x+q}/2}\rho_A 
    \ket{{\bf x-q}/2}
    e^{\im {\bf p\cdot q}}.
\end{align}
In general, the Wigner function of the bosonic Gaussian state can be written in terms of its covariance matrix $\Gamma$ and the mean vector $\boldsymbol{s}$ as 
\begin{align}
    W({\bf x,p})
    =
    \frac{\exp(-\frac{1}{2}({\bf r}-\boldsymbol{s})^T\Gamma^{-1}({\bf r}-\boldsymbol{s}))}
    {(2\pi)^{V_A}\sqrt{\det(\Gamma)}}, 
\end{align}
where ${\bf r}$ is the $2V_A$-dimensional vector whose entries are given by ${\bf r}_{\bf i}=({\rm x}_{\bf i},{\rm p}_{\bf i})^T$.
Note that this is different from the operators $\boldsymbol{r}_{\bf i}=(x_{\bf i},p_{\bf i})^T$ in the main text. 
In the same way, the Wigner function of $\rho_{A,\chi}$ can be obtained as 
\begin{align}
    W_\chi({\bf x,p})
    =
    \frac{\exp(-\frac{1}{2}({\bf r}-\boldsymbol{s}_\chi)^T\Gamma^{-1}_\chi({\bf r}-\boldsymbol{s}_\chi))}
    {(2\pi)^{V_A}\sqrt{\det(\Gamma_\chi)}}
    \label{eq:W_chi}
    .
\end{align}
\par 
In the Wigner function formalism, the expectation value of an operator $O$ in the state $\rho_A$ can be written as 
\begin{align}
    \Tr_{}(\rho_AO)
    =
    \int_{\mathbb{R}^{2V_A}}
    {\rm d}{\bf x}
    {\rm d}{\bf p}\,
    W({\bf x,p})
    F_O({\bf x,p}), 
    \label{eq:expectation_Wigner}
\end{align}
where 
\begin{align}
    F_O({\bf x,p})
    = 
    \int_{\mathbb{R}^{V_A}}
    {\rm d}{\bf q}
    \bra{{\bf x+q}/2} O
    \ket{{\bf x-q}/2}
    e^{\im {\bf p\cdot q}}.
\end{align}
From Eqs.~\eqref{eq:W_chi} and~\eqref{eq:expectation_Wigner}, the charged moment $Z_\chi=\Tr_{}(\rho_A\rho_{A,\chi})$ can be written as 
\begin{align}
    Z_\chi
    &= 
    (2\pi)^{V_A}
    \int_{\mathbb{R}^{2V_A}}
    {\rm d}{\bf x}
    {\rm d}{\bf p}\,
    W({\bf x,p})
    W_\chi({\bf x,p})
\end{align}
and, therefore,
\begin{align}
    Z_\chi&=
    (2\pi)^{V_A}
    \frac{\exp\left[
    -\frac{1}{2}
    (\boldsymbol{s}-\boldsymbol{s}_\chi)^T
    [\Gamma+\Gamma_\chi]^{-1}
    (\boldsymbol{s}-\boldsymbol{s}_\chi)\right]}
    {\sqrt{\det(\Gamma\Gamma_\chi)}}
    \nonumber\\
    &\times
    \int_{\mathbb{R}^{2V_A}}
    {\rm d}{\bf x}{\rm d}{\bf p}\,
    \exp\left[-\frac{1}{2}({\bf r}-\tilde{\boldsymbol{s}}_\chi)^T
    \tilde \Gamma_\chi^{-1} 
    ({\bf r}-\tilde{\boldsymbol{s}}_\chi)\right], 
    \label{eq:WW}
\end{align}
where 
\begin{gather}
    \tilde \Gamma_\chi (t)
    = 
    (\Gamma(t)^{-1}+\Gamma_\chi(t)^{-1})^{-1},
    \\
    \tilde{\boldsymbol{s}}_\chi(t)
    =
    \tilde \Gamma_\chi(t)(\Gamma(t)^{-1}\boldsymbol{s}+\Gamma_\chi(t)^{-1}\boldsymbol{s}_\chi). 
\end{gather}
Calculating the Gaussian integral in Eq.~\eqref{eq:WW}, we arrive at Eq.~\eqref{eq:charged moment exact}.

\section{Derivation of Eqs.~\eqref{eq:A(t) analytic} and \eqref{eq:f(t) analytic}}\label{apdx:mult_st_app}

Here we employ the results of Appendix~\ref{apdx:stationary_phase} to derive Eqs.~\eqref{eq:A(t) analytic} and~\eqref{eq:f(t) analytic}. 

{\it Eq.~\eqref{eq:A(t) analytic}.---} We expand Eq.~\eqref{eq:A} in terms of $\Lambda_{k_y,\chi}(t)$ as 
\begin{align}
    A_\chi(t)
    =
    \frac{1}{2V_A}
    \sum_{k_y}
    \sum_{m=1}^\infty 
    \qty(
    \frac{\Tr_{}(\tilde \Lambda_{k_y,\chi}^m)}{m}
    -
    \frac{\Tr_{}(\tilde \Lambda_{k_y,0}^m)}{m}). 
    \label{eq:A(t) expand}
\end{align}
where $\tilde \Lambda_{k_y,\chi}(t)=I-\Lambda_{k_y,\chi}(t)/\gamma$ and $\gamma$ is a positive constant larger than the largest eigenvalue of $\Lambda_{k_y,\chi}$. 
We introduced $\gamma$ to ensure $\lim_{m\to\infty}\tilde \Lambda_{k_y,\chi}^m=0$, but it should disappear in the final result because it is an arbitrary parameter. We can calculate the moments of $\tilde{\Lambda}_{k_y, \chi}$ applying directly Eq.~\eqref{eq:Tr(T^m)_final}. The symbol of $\tilde{\Lambda}_{k_y, \chi}$ is $I- \lambda_{\bf k}(t)/\gamma$, where $\lambda_{\bf k}(t)$ was defined in Eq.~\eqref{eq:lambda_k}. Identifying $c_{k_x}$ and $d_{k_x}$ with $1-\cos(\theta_{\bf k})/\gamma$ and $\sinh(\theta_{\bf k})\cos(\chi)/\gamma$, respectively, Eq.~\eqref{eq:Tr(T^m)_final} reduces to
\begin{widetext}
\begin{multline}
    \Tr_{}(\tilde \Lambda_{k_y,\chi}(t)^m)
    \simeq 
    \ell\int_{-\pi}^\pi 
    \frac{{\rm d}k_x}{2\pi}
    x_{k_x}(\zeta)
    \Bigg[\qty(1-\frac{\cosh(\theta_{\bf k})}{\gamma}-\frac{\sinh(\theta_{\bf k})\cos(\chi)}{\gamma})^m
    +
    \qty(1-
    \frac{\cosh(\theta_{\bf k})}{\gamma}
    +\frac{\sinh(\theta_{\bf k})\cos(\chi)}{\gamma})^m\Bigg]
    \\
    +
    2
    \int_{-\pi}^\pi 
    \frac{{\rm d}k_x}{2\pi}
    \min(\ell,2t|v_{k_x}|)
    \qty(1-\frac{\cosh(\theta_{\bf })}{\gamma})^m
    .
    \label{eq:Lambda tilde moment}
\end{multline}
\end{widetext}
Inserting Eq.~\eqref{eq:Lambda tilde moment} into Eq.~\eqref{eq:A(t) expand} and performing the summation over $m$, we finally obtain Eq.~\eqref{eq:A(t) analytic}.

{\it Eq.~\eqref{eq:f(t) analytic}.---} Let us start from Eq.~\eqref{eq:f_k(t)} and take first the term $\log(\Gamma(t)+\Gamma(\infty))$. We expand it in terms of $\Phi_{k_y}(t):=I-\Gamma_{k_y}(t)-\Gamma_{k_y}(\infty)$ as 
\begin{align}
    \log \det(\Gamma_{k_y}(t)+\Gamma_{k_y}(\infty))&=\Tr_{}\log(I-\Phi_{k_y}(t))
    \nonumber\\
    &=
    -\sum_{m=1}^\infty 
    \frac{\Tr_{}([\Phi_{k_y}(t)]^m)}{m}.
    \label{eq:first term expand}
\end{align}
In the thermodynamic limit $L_x\to\infty$, $\Phi_{k_y}(t)$ becomes a block-Toeplitz matrix generated by the symbol 
\begin{align}
    \phi_{\bf k}(t)
    &=
    I-g_{\bf k}(t)-g_{\bf k}(\infty)
    \nonumber\\
    &= 
    (1-\cosh\theta_{\bf k}) I 
    - 
    \frac{\sinh\theta_{\bf k}}{2}\sigma_z e^{-2\im t \xi_{\bf k}\sigma_y}. 
\end{align}
Identifying $c_{k_x}$ with $1-\cosh(\theta_{\bf k})$ and $d_{k_x}$ with $-\sinh(\theta_{\mathbf{k}})/2$ in Eq.~\eqref{eq:Tr(T^m)_final}, we find
\begin{align}
    &\Tr_{}([\Phi_{k_y}(t)]^m)
    \nonumber
    \\
    &\simeq 
    \ell 
    \int_{-\pi}^\pi
    \frac{{\rm d}k_x}{2\pi}
    x_{k_x}(\zeta)
    [(1-\cosh(\theta_{\bf k})+\sinh(\theta_{\bf k})/2)^m
    \nonumber
    \\
    &+
    (1-\cosh(\theta_{\bf k})-\sinh(\theta_{\bf k})/2)^m]
    \nonumber
    \\
    &+
    2\ell
    \int_{-\pi}^\pi
    \frac{{\rm d}k_x}{2\pi}
    \min(1,2\zeta|v_{k_x}|)
   (-1)^m\cosh(\theta_{\bf k})^m.
    \label{eq:Phi moment asymptotic}
\end{align}
Inserting Eq.~\eqref{eq:Phi moment asymptotic} into Eq.~\eqref{eq:first term expand} and performing the summation over $m$, we obtain 
\begin{align}
    &\log
    \det(\Gamma_{k_y}(t)+\Gamma_{k_y}(\infty))
    \\
    &\simeq 
   \ell\int_{-\pi}^\pi
    \frac{{\rm d}k_x}{2\pi}
    x_{k_x}(\zeta)
    \log(1+\frac{3}{4}\sinh^2(\theta_{\bf k}))
    \nonumber
    \\
    &\quad 
    +
    2\ell
    \int_{-\pi}^\pi
    \frac{{\rm d}k_x}{2\pi}
    \min(1,2\zeta|v_{k_x}|)
    \log(\cosh(\theta_{\bf k})). 
    \label{eq:f_k first term}
\end{align}
\par 
We move now on to study the term $\log \det C_{k_y}^+(t)$ of Eq.~\eqref{eq:f_k(t)}. The matrix $C_{k_y}^+(t)$ is a product of block-Toeplitz matrices and, in general, it is not a block-Toeplitz matrix. 
However, for large $\ell$, the product of block-Toeplitz matrices can be approximated as the block-Toeplitz matrix generated by the product of the symbols of the factors, see e.g.~\cite{amvc-23}. 
This allows us to approximate $C_{k_y}^+(t)$ by the block-Toeplitz matrix generated by 
\begin{align}
    c_{\bf k}^+(t)
    &= 
    -2\sigma_y 
    (g_{\bf k}(t)+g_{\bf k}(\infty))^{-1}
    (\sigma_y/4+g_{\bf k}(\infty)\sigma_yg_{\bf k}(t))
    \nonumber\\
    &=
    \cosh\theta_{\bf k}I
    +\sinh\theta_{\bf k}\sigma_z e^{-2\im t \xi_{\bf k}\sigma_y}.
    \label{eq:c_k}
\end{align}
Taking this into account, the asymptotics of $\log\det C_{n_y}^+(t)$  can be calculated by expressing it in terms of the moments of $I-C_{k_y}^+$ and evaluating $\Tr((I-C_{k_y}^+)^m)$ using Eq.~\eqref{eq:Tr(T^m)_final}. The result is 
\begin{align}
    \log \det C_{n_y}^+(t)
    \simeq 
    \int_{-\pi}^\pi
    \frac{{\rm d}k_x}{2\pi}
    \min(1,2\zeta|v_{k_x}|)\log(\cosh(\theta_{\bf k})).
    \label{eq:f_k second term}
\end{align}
\par 
We finally calculate the term $\log\det C_{k_y}^-(t)$ Eq.~\eqref{eq:f_k(t)}. 
As we have done for the matrix $C_{k_y}^+(t)$, we approximate $C_{k_y}^-(t)$ as the block Toeplitz matrix generated by 
\begin{align}
    c_{\bf k}^-(t)
    &= 
    \sqrt{I-(c_{\bf k}\sigma_y)^{-2}/8}+I
    =I.
\end{align}
The above equation implies that $C_{n_y}^-(t)\simeq I$  for $\ell\gg1$ because the block-Toeplitz matrix with symbol the identity is the identity. 
We thus obtain 
\begin{align}
   \log \det C_{n_y}^-(t)\simeq 0.
    \label{eq:f_k third term}
\end{align}
Applying the results in Eqs.~\eqref{eq:f_k first term}, \eqref{eq:f_k second term}, and \eqref{eq:f_k third term} in Eq.~\eqref{eq:f_k(t)}, we finally obtain Eq.~\eqref{eq:f(t) analytic}.

\section{Derivation of Eq.~\eqref{eq:B(t) analytic}}\label{apdx:B_chi}

In this Appendix, we obtain the asymptotic expression~\eqref{eq:B(t) analytic} for $B_\chi(t)$. To this end, we rewrite Eq.~\eqref{eq:B} as 
\begin{align}
    B_\chi(t)
    = 
    \frac{4n_0\sin(\frac{\chi}{2})^2}{\ell \gamma}
    \boldsymbol{u}^T[I -\tilde \Lambda_{0,\chi}(t)]^{-1}\boldsymbol{u}. 
\end{align}
Since $\lim_{m\to\infty}[\tilde \Lambda_{k_y,\chi}(t)]^m=0$, we can perform the Neumann expansion of $[I -\tilde \Lambda_{0,\chi}(t)]^{-1}$ in the above equation and obtain 
\begin{align}
    B_\chi(t)
    = 
    \frac{4n_0\sin(\frac{\chi}{2})^2}{\ell \gamma}
    \sum_{m=0}^\infty 
    \boldsymbol{u}^T [\tilde \Lambda_{0,\chi}(t)]^m \boldsymbol{u}. 
    \label{eq:B(t) expand}
\end{align}
Once again, the asymptotic form of $\eta_m:=\boldsymbol{u}^T [\tilde \Lambda_{0,\chi}(t)]^m \boldsymbol{u}$ in the ballistic limit can be derived by applying the multidimensional stationary phase method. In the thermodynamic limit, $\tilde{\Lambda}_{k_y,\chi}$ is the block-Toeplitz matrix with  symbol $\tilde \lambda_{\bf k}=I-\lambda_{\bf k}/\gamma$. Therefore, using the identity~\eqref{eq:sum_to_integral}, $\eta_m$ can be written as 
\begin{multline}
    \eta_m
    =
    \qty(\frac{\ell}{2})^{m+1}
    \int \limits_{[-1,1]^{m+1}}
    \prod_{i=0}^m {\rm d}u_i
    \int \limits_{[-\pi,\pi]^m}
    \prod_{i=1}^m 
    \frac{{\rm d}k_i}{2\pi}
    \nonumber\\
    \times 
    C(\boldsymbol{k})
    G(\boldsymbol{k})
    e^{-\frac{\im}{2}\ell \sum_{i=1}^m k_i(u_{i-1}-u_i)},
    \label{eq:uLamu_interal_1}
\end{multline}
where $C(\boldsymbol{k})$ was defined in~\eqref{eq:C_k} and 
\begin{align}
    G(\boldsymbol{k})
    &=
    (0,1)
    \qty(\prod_{i=1}^m \tilde{\lambda}_{k_i})
    \mqty(0\\1).
\end{align}
Using the identity, 
\begin{align}
    1=
    \int_{-\pi}^\pi
    {\rm d}k_0
    e^{\frac{\im}{2}\ell k_0(i_0-i_m)}\delta(k_0), 
\end{align}
Equation~\eqref{eq:uLamu_interal_1} can be written as 
\begin{multline}\label{eq:uLu_integral_3}
    \eta_m=
    2\pi 
    \qty(\frac{\ell}{2})^{m+1}
    \int \limits_{[-1,1]^{m+1}}
    \prod_{i=0}^m {\rm d}u_i
    \int \limits_{[-\pi,\pi]^{m+1}}
    \prod_{i=0}^m 
    \frac{{\rm d}k_i}{2\pi}
   \\
    \times
    C(\boldsymbol{k})
    G(\boldsymbol{k})
    e^{\frac{\im}{2}\ell \sum_{i=0}^m k_i(u_{i}-u_{i-1})}
    \delta(k_0).
\end{multline}
In the ballistic limit $\ell\gg1$, the delta function $\delta(k_0)$ can be expressed as 
\begin{align}
    \delta(k_0)
    \simeq 
    d_\ell(k_0),~
    d_\ell(k_0)
    =
    \frac{\sin^2(\frac{\ell k_0}{2})}{2\pi \ell  \sin^2(\frac{k_0}{2})}.
    \label{eq:delta=d}
\end{align}
Inserting Eq.~\eqref{eq:delta=d} into Eq.~\eqref{eq:uLu_integral_3}, we obtain 
\begin{multline}
    \eta_m
    \simeq 
    2\pi 
    \qty(\frac{\ell}{2})^{m+1}
    \int \limits_{[-1,1]^{m+1}}
    \prod_{i=0}^m {\rm d}u_i
    \int \limits_{[-\pi,\pi]^{m+1}}
    \prod_{i=0}^m 
    \frac{{\rm d}k_i}{2\pi}
    \\
    \times 
    C(\boldsymbol{k})
    G(\boldsymbol{k})
    e^{\frac{\im}{2}\ell \sum_{i=0}^m k_i(u_{i}-u_{i-1})}
    d_\ell(k_0).
    \label{eq:uLu_integral_4}
\end{multline}
Introducing the new variables 
\begin{align}
    z_i
    =
    \begin{dcases}
        u_0 &i=0,\\
        u_i-u_{i-1}&i=1,...,m,
    \end{dcases}
\end{align}
and, taking the integral over $z_0$, 
Eq.~\eqref{eq:uLu_integral_4} can be written as 
\begin{multline}
    \eta_m\simeq 
    2\pi 
    \qty(\frac{\ell}{2})^{m+1}
    \int \limits_{\mathbb{R}^m}
    \prod_{i=1}^m {\rm d}z_i
    \int \limits_{[-\pi,\pi]^{m+1}}
    \prod_{i=0}^m 
    \frac{{\rm d}k_i}{2\pi}
    \\ 
    \times 
    \nu(\boldsymbol{z})
    C(\boldsymbol{k})
    G(\boldsymbol{k})
    e^{\frac{\im}{2}\ell \sum_{i=1}^m z_i(k_i-k_0)}
    d_\ell(k_0), 
\end{multline}
where 
\begin{align}
    \nu(\boldsymbol{z})
    =
    \max\qty[
    0,
    2-\max_{j\in[0,m]}
    \qty(\sum_{i=1}^jz_i)
    +
    \min_{j\in[0,m]}
    \qty(\sum_{i=1}^jz_i)
    ]. 
\end{align}
 The explicit calculation of the product of matrices $\tilde{\lambda}_{k_y}$ in $G(\boldsymbol{k})$ gives
 \begin{widetext}
\begin{align}
    &G(\boldsymbol{k})=
    \sum_{p=0}^m 
    \sum_{1\leq j_1<...<j_p\leq m}
    \cos(2t\sum_{i=1}^p (-1)^i \xi_{k_{j_i}})
    \qty[
    \prod_{j=1}^p
    \qty(\frac{\sinh(\theta_{k_{j_i}})\cos(\chi)}{\gamma})]
    \qty[
    \prod_{i\notin\{j_1,...,j_p\}}
    \qty(1-\frac{\cosh(\theta_{k_i})}{\gamma})]
    .
    \label{eq:uLu_integral_2}
\end{align}
\end{widetext}
 Using this expression and taking into account that the integral is invariant under the permutations $z_{j_i}\leftrightarrow z_{i-1}$ and $k_{j_i}\leftrightarrow k_{i-1}$, we obtain 
\begin{align}
    \eta_m\simeq 
    \qty(\frac{\ell}{2})^{m+1}
    \int_{-\pi}^\pi
    {\rm d}k_0
    d_\ell(k_0)
    \sum_{p=0}^m 
    \mqty(m\\p)
    \Re[\tilde{\mathcal{I}}_p(k_0)]. 
    \label{eq:u T u int}
\end{align}
with 
\begin{multline}
    \tilde{\mathcal{I}}_p(k_0)
    =
    \int \limits_{\mathbb{R}^{m}}
    \prod_{i=1}^m 
    {\rm d}z_i
    \int \limits_{[-\pi,\pi]^m}
    \prod_{i=1}^m
    \frac{{\rm d}k_i}{2\pi}
    D(\boldsymbol{k},\boldsymbol{z})
    e^{\im \ell \varphi(\boldsymbol{k},\boldsymbol{z})}
    ,
    \label{eq:tilde_I_p}
\end{multline}
where
\begin{align}
    &D(\boldsymbol{k},\boldsymbol{z})
    =
    \nu(\boldsymbol{z})
    C(\boldsymbol{k})
    \nonumber\\
    &\times 
    \qty[
    \prod_{i=p}^{m-1}
    \qty(1-\frac{\cosh(\theta_{k_i})}{\gamma})
    ]
    \qty[
    \prod_{i=0}^{p-1}
    \qty(\frac{\sinh(\theta_{k_i})\cos(\chi)}{\gamma})
    ],
\end{align}
and
\begin{align}
    &\varphi(\boldsymbol{k},\boldsymbol{z})
    =
    \frac{1}{2}
    \sum_{i=1}^m
    z_i(k_i-k_0)
    -
    2\zeta\sum_{i=0}^{p-1}(-1)^i\xi_{k_i}.
\end{align}
We now evaluate $\tilde{\cal I}_p(k_0)$ using the stationary phase approximation. 
The stationary points in the present case are given by $\nabla \varphi(\bar{\boldsymbol{k}},\bar{\boldsymbol{z}})=0$ and read
\begin{align}
    \bar{k}_i = k_0,~
    \bar{z}_i 
    =
    \begin{cases}
        4\zeta(-1)^i v_{k_0},&i=1,...,p-1,\\
        0,&i=p,...,m.
    \end{cases}
    \label{eq:stationary points for tilde_I_p}
\end{align}
At the stationary points~\eqref{eq:stationary points for tilde_I_p}, we have 
\begin{align}
    |{\rm det (H}(\varphi(\bar{\boldsymbol{k}},\bar{\boldsymbol{z}})))|
    &=
    4^{1-m},~
    {\rm sgn (H}(\varphi(\bar{\boldsymbol{k}},\bar{\boldsymbol{z}})))=0,
    \label{eq:Hess}
\end{align}
\begin{align}
    D(\bar{\boldsymbol{k}},\bar{\boldsymbol{z}})
    &=
    \qty(1-\frac{\cosh(\theta_{k_x})}{\gamma})^{m-p}
    \qty(\frac{\sinh(\theta_{k_x})\cos(\chi)}{\gamma})^p 
    \nonumber\\
    &\quad \times 
    \begin{dcases}
        2 &p=0,1,\\
        2x_{k_0}(\zeta)&p=2,3,...,m.
    \end{dcases}
    \label{eq:D}
\end{align}
and
\begin{align}
    \varphi(\bar{\boldsymbol{k}},\bar{\boldsymbol{z}})
    &=
    \begin{cases}
        0&p\quad {\rm even},\\
        2\zeta\xi_{k_0}&p\quad {\rm odd}.
    \end{cases}
    \label{eq:verphi}
\end{align}

Applying the formula~\eqref{eq:formula_stationary_phsae} to Eq.~\eqref{eq:tilde_I_p} and using Eqs.~\eqref{eq:Hess},~\eqref{eq:verphi}, and~\eqref{eq:D}, we obtain 
\begin{align}
    \tilde{\mathcal{I}}_{p}(k_x)
    &\simeq 
    \qty(\frac{4\pi}{\ell})^m 
    \qty(1-\frac{\cosh(\theta_{k_x})}{\gamma})^{m-p}
    \qty(\frac{\sinh(\theta_{k_x})\cos(\chi)}{\gamma})^p 
    \nonumber\\
    &
    \quad 
    \times
    \begin{dcases}
        2&p=0,\\
        2e^{-2\im t \xi_{k_x}} &p=1,\\
        2x_{k_x}(\zeta)&p=2,4,6,...,\\
        2x_{k_x}(\zeta)e^{-2\im t \xi_{k_x}}&p=3,5,7,....
    \end{dcases}
    \label{eq:tilde_I_stationary}
\end{align}
Here we have rewritten $k_0$ as $k_x$. 
Substituting Eq.~\eqref{eq:tilde_I_stationary} into Eq.~\eqref{eq:u T u int} and performing the sum over $p$, we arrive at 
\begin{widetext}
\begin{multline}
   \boldsymbol{u}^T [\tilde \Lambda_{0,\chi}(t)]^m \boldsymbol{u}
    \simeq 
    \ell 
    \int_{-\pi}^\pi 
    {\rm d}k_x 
    d_\ell(k_x)
    x_{k_x}(\zeta)
   \\ \times \Bigg[
    \qty(1-\frac{\cosh(\theta_{k_x})}{\gamma}
    +\frac{\sinh(\theta_{k_x})\cos(\chi)}{\gamma}
    )^m \cos^2(t\xi_{k_x})
    +
    \qty(1-\frac{\cosh(\theta_{k_x})}{\gamma}
    -\frac{\sinh(\theta_{k_x})\cos(\chi)}{\gamma}
    )^m \sin^2(t\xi_{k_x})
    \Bigg]
    \\+
    \ell 
    \int_{-\pi}^\pi 
    {\rm d}k_x 
    d_\ell(k_x)
    [1-x_{k_x}(\zeta)]
    \Bigg[
    \qty(1-\frac{\cosh(\theta_{\bf k})}{\gamma})^m
    +m
    \qty(1-\frac{\cosh(\theta_{\bf k})}{\gamma})^{m-1}
    \frac{\sinh(\theta_{k_x})\cos(\chi)}{\gamma}\cos(2t\xi_{k_x})
    \Bigg], 
    \label{eq:u Lam u}
\end{multline}
\end{widetext}
where $\xi_{k_x}=\xi_{(k_x,0)}$. Inserting this expression in Eq.~\eqref{eq:B(t) expand} and performing the summation over $m$, we finally find Eq.~\eqref{eq:B(t) analytic} of the main text.

\end{document}